\documentclass[10pt,twocolumn,showpacs,nofootinbib,aps,prd]{revtex4-2}
\usepackage{latexsym}
\usepackage{graphicx}
\usepackage{dcolumn}
\usepackage{bm}
\usepackage{xcolor}
\usepackage{empheq}
\usepackage{float}
\usepackage{enumitem}
\usepackage{amssymb,amsmath}
\usepackage{slashed}
\usepackage{hyperref}

\usepackage{CJK} 

\usepackage{latexsym,amsfonts,amsmath,amssymb}
\usepackage{bbm}
\usepackage{hyperref}
\usepackage{empheq}
\usepackage{graphicx}
\usepackage{color}
\usepackage{caption}
\usepackage{subcaption}
\usepackage[normalem]{ulem}
\usepackage{comment}
\usepackage{url}
\usepackage{slashed}
\usepackage{multirow}
\usepackage{pifont}
\usepackage{amsxtra,graphics,epsfig,bm,tikz,xfrac}
\usetikzlibrary{decorations.pathmorphing}
\usetikzlibrary{decorations.markings}
\usetikzlibrary{arrows, decorations.markings, calc, fadings, decorations.pathreplacing, patterns, decorations.pathmorphing, positioning}

\usetikzlibrary{positioning,shapes}
\usetikzlibrary{chains}
\usetikzlibrary{arrows,fit,decorations.pathreplacing}
\tikzstyle{every picture}+=[remember picture]
\tikzstyle{na} = [baseline=-.5ex]

\usepackage{float}
\usepackage[english]{babel}
\usepackage[utf8]{inputenc}
\usepackage[colorinlistoftodos, color=green!40, prependcaption]{todonotes}
\usepackage{physics}
\usepackage{xcolor}
\usepackage{adjustbox}
\usepackage{placeins}
\usepackage[T1]{fontenc}
\usepackage[normalem]{ulem} 

\usepackage{float}
\usepackage[english]{babel}
\usepackage[utf8]{inputenc}
\usepackage[colorinlistoftodos, color=green!40, prependcaption]{todonotes}
\usepackage{physics}
\usepackage{xcolor}
\usepackage{adjustbox}
\usepackage{placeins}
\usepackage[T1]{fontenc}
\usepackage[normalem]{ulem} 

\newcommand{\be}{\begin{equation}}
\newcommand{\ee}{\end{equation} }
\newcommand{\beqa}{\begin{eqnarray} }
\newcommand{\eeqa}{\end{eqnarray} }
\newcommand{\ba}{\begin{array}}
\newcommand{\ea}{\end{array}}

\def\ads2{AdS$_2$}

\begin{document}
\begin{CJK}{UTF8}{mj}
\title{Stability and topological nature of charged Gauss-Bonnet AdS black holes in five dimensions}

\author{Imtak Jeon} 
\email{imtakjeon@gmail.com} 
\affiliation{
School of Science, Huzhou University, Huzhou 313000, Zhejiang, China}
\affiliation{
 Asia Pacific Center for Theoretical Physics, Postech, Pohang 37673, Korea }
\affiliation{
 Department of Physics, Postech, Pohang 37673, Korea 
}

\author{Bum-Hoon Lee}
\email{bhl@sogang.ac.kr}
\affiliation{
Center for Quantum Spacetime, Sogang University, Seoul 04107, Republic of Korea }
\affiliation{
 Department of Physics, Sogang University, Seoul 04107, Republic of Korea  }

\author{Wonwoo Lee} 
\email{warrior@sogang.ac.kr}
\affiliation{
Center for Quantum Spacetime, Sogang University, Seoul 04107, Republic of Korea }

\author{Madhu Mishra} 
\email{madhu.mishra@apctp.org}
\affiliation{
 Asia Pacific Center for Theoretical Physics, Postech, Pohang 37673, Korea }

\begin{abstract}
	\noindent
We investigate the thermodynamics of Reissner-Nordstr\"om Gauss-Bonnet (RN-GB) black holes in anti-de Sitter (AdS) space with three horizon geometries ($k = +1, 0, -1$) within the grand canonical ensemble. Using the recently developed topological approach to black hole thermodynamics, inspired by Duan's $\phi$-mapping theory, we analyze the black holes by treating both critical points in the phase diagram and black hole solutions as defects in the thermodynamic parameter space. Our results show that the Gauss-Bonnet coupling significantly alters the topological classification of RN-GB AdS black holes, distinguishing them from their RN AdS counterparts in the grand canonical ensemble, while aligning with their canonical ensemble counterparts. Complementary analyses of local stability using specific heat validate the implication of topological analysis. Furthermore, an evaluation of global stability via Gibbs free energy provides a comprehensive understanding on system's phase structure. Notably, for $k=+1$, topological analysis suggests liquid-gas type phase transitions, whereas global analysis favors Hawking-Page transitions. For $k=-1$, topology indicates a single stable black hole branch, yet the global analysis reveals the presence of Hawking-Page transitions.

\end{abstract}
\maketitle
\end{CJK}

\section{Introduction} \label{sec:introduction}

Black hole thermodynamics is one of the most fascinating subjects in theoretical physics. The thermal nature of black holes, characterized by  their temperature and thermal entropy \cite{Bekenstein:1973ur, Hawking:1975vcx}, provides important theoretical guidance for exploring the quantum aspects of gravitational interactions. As thermal objects, understanding their stability and phase structures has been an area of active research for several decades. 

The static and spherically symmetric Schwarzschild black hole is unstable due to its negative specific heat. This negativity means that as the black hole loses energy (e.g., through Hawking radiation), its temperature increases, causing it to lose energy even faster. This self-reinforcing cycle leads to a runaway process, making the black hole unstable. To address this issue, researchers have employed mathematical treatments, such as placing the black hole in a large box \cite{York:1986it}. A modern approach involves placing black holes in anti-de-Sitter (AdS) spacetime, which ensures thermal stability for sufficiently large black holes. This setup features a transition between large black holes and thermal gas in AdS space at low temperature, known as the Hawking-Page transition~\cite{Hawking:1982dh}. 

Subsequently, more complex black hole systems have been studied, including charged black holes (Reissner-Nordstr\"{o}m, RN) and black holes with higher-derivative corrections such as Gauss-Bonnet (GB) couplings in various dimensional AdS spacetimes \cite{Braden:1990hw,Chamblin:1999tk,Chamblin:1999hg,Dey:2006ds,Cai:2013qga,Dey:2007vt,Anninos:2008sj,Zou:2014mha,Abdusattar:2023fdm,Konoplya:2017ymp,Konoplya:2017zwo}. In canonical ensemble, where the black hole charge is fixed, an additional stable black hole phase emerges at low temperatures, resulting in phase transition between small and large black holes, with or without GB corrections.  This phase transition is analogous to the liquid-gas phase transition in the van der Waals system, and this analogy has been enhanced using the `extended' phase space, where the cosmological constant is interpreted as thermodynamic pressure  \cite{Kastor:2009wy,Dolan:2010ha, Dolan:2011xt,Kubiznak:2012wp,Wei:2014hba}. 
In the grand canonical ensemble, where the electric potential is fixed, allowing emission and absorption of the charged particles,  no stable small black hole phase exists, and the liquid-gas type phase transition does not occur for spherical RN AdS black holes \cite{Chamblin:1999tk,Chamblin:1999hg,Braden:1990hw} without GB coupling. However, the presence of GB coupling introduces a new stable small black hole phase, enabling liquid-gas type phase transitions 
\cite{Dey:2007vt,Anninos:2008sj,Zou:2014mha}.

Recently, the study of black hole thermodynamics has gained fresh insights by incorporating the concept of topology. Using Duan's topological $\phi$-mapping theory \cite{Duan:1984ws}, critical points in the phase diagram are treated as defects \cite{Wei:2021vdx}, or black hole solutions themselves are considered defects \cite{Wei:2022dzw} in the thermodynamic parameter space. Assigning topological charges to these defects has illuminated the properties of the critical points and the local stability of different black hole solution branches, offering a universal classification of black hole systems \cite{Yerra:2022alz,Yerra:2022eov,Liu:2022aqt,Wu:2022whe,Fang:2022rsb,Wu:2023sue,Wu:2023meo,Wu:2024rmv,Zhu:2024jhw,Wu:2024txe,Chen:2024atr,Hazarika:2024xar,Liu:2024fvq,Zhu:2024zcl,Rathi:2024ycw,Wei:2024gfz,Wu:2024asq,Wang:2024zlq}.

In \cite{Liu:2022aqt} the local and global topological properties of charged Gauss-Bonnet black
holes in AdS space in the canonical ensemble were investigated, revealing that the GB coupling does not alter the topological number in five dimensions. Additionally, this topological number remains unaffected by the geometry of black hole horizon, whether  spherical, planar, or hyperbolic. Motivated by these findings, this paper focuses on  thermodynamics of the five dimensional RN-GB AdS black holes in the grand canonical ensemble, exploring three different horizon geometries: spherical $(k = +1)$, planar $(k = 0)$, and hyperbolic $(k = -1)$. 
For our topological analysis, we follow the approach of \cite{Wei:2021vdx,Wei:2022dzw} to determine the properties of critical points, providing their novel classifications  and the local stability of black hole solutions. In the grand canonical ensemble, the GB coupling plays a critical role, leading to a topological number for RN-GB AdS black holes that differs from that of RN AdS black holes and aligns with the topological number of RN-AdS or RN-GB AdS black holes in the canonical ensemble.

As a complementary study, we examine local stability using specific heat and analyze phase transitions using global stability provided by the Gibbs free energy. This analysis reaffirms the local stability predicted by topological numbers for each solution and provides insights into global phase structures, which are not addressed by topological studies. While the topological analysis indicates that the RN-GB AdS black holes in the grand canonical ensemble can exhibit liquid-gas type phase transitions for spherical horizon geometry $(k = 1)$, the global analysis reveals a preference for Hawking-Page type phase transitions. For the hyperbolic horizon geometry $(k = -1)$, topology indicates a single stable black hole branch, but global analysis shows Hawking-Page transitions.

The paper is structured as follows: In Section II, we review Duan's $\phi$-mapping theory and its two applications - treating critical points in the phase diagram as defects and treating black hole solutions as defects. In Section III, we present the setup for charged GB AdS black holes in five dimensions, along with thermodynamic variables and ranges of the parameters. In Section IV, we analyze the topological nature of black holes by examining critical points and black hole solutions. In Section V, we provide complementary analyses using specific heat to confirm the local stability predicted by topology and further explore global phase structures using Gibbs free energy, which are not captured in the topology study. Finally, we conclude in Section VI.

\section{Topological current $\phi$-mapping theory}\label{PhimapTheory}
In this section, we review the topological analysis using Duan's topological current $\phi$-mapping theory~\cite{Duan:1984ws} and its two application to the black hole thermodynamics following~\cite{Wei:2021vdx,Wei:2022dzw}.

Consider a two component vector order parameter field $\phi=(\phi^{1},\phi^2)$ defined on a certain parameter space. The key concept in the study of topology is the identification of defects, which are singular points of the vector field $\phi^a(z_i)=0$, where $z_i$ denotes the location of these zero points. The topological current associated with such defects is given by:
\begin{align}
	j^\mu = \frac{1}{2 \pi} \epsilon^{\mu\nu\rho} \epsilon_{ab} \partial_\nu n^a \partial_\rho n^b\,,\quad (\mu=0,1,2),
\end{align}
where $n^a = \frac{\phi^a}{||\phi^a||}\,(a=1,2)$ is the normalized unit vector.
 A distinctive feature of this topological currents is that it is conserved independently of field equations and is non-zero only at the defect point $\phi^a = 0$. The topological charge, defined as the integral of the current density $j^0$ over the parameter space $\Sigma$, is 
 \begin{equation}\label{topological_charge}
 	Q_t=\int_\Sigma j^0 d^2x=\sum_{i=1} ^N  w_i\,,
 \end{equation}
where $w_i$ represents the winding number of the vector field $\phi$ around its $i$-th zero point $z_i$. The topological charge $Q_t$, being an integer, is the sum of these winding numbers.

The $\phi$-mapping theory has been applied to black hole thermodynamics in two primary ways:  One is identifying critical points in the phase diagram as the defects and assigning topological charges, and the other is interpreting black hole solutions as the defects, enabling their classification and stability analysis. We shall outline these approaches below.
\\

\noindent
\textit{Critical points as defects:} To apply the theory to ciritical points, we consider temperature $T(r_+,P,x)$, as a function of the black hole's outer horizon $r_+$, pressure $P$ and the other variables denoted by $x$.  Then, using the condition $\left(\frac{\partial T}{\partial r_{+}}\right)_{P,x
}=0$, we eliminate the pressure $P$ dependence and define a Duan's potential as \cite{Wei:2021vdx}:
\begin{equation}\label{duan_poten}
\Phi=\frac{1}{\sin \theta} T(r_+,z^i)\,,
\end{equation}
where an additional auxiliary coordinate $\theta$ is introduced to facilitate the construction of a  two-dimensional vector field in the $r_+ - \theta$ plane, and the factor $1/\sin\theta$ is multiplied for the ease of analysis. The corresponding vector field is defined as 
\begin{equation}
\phi^{r_+}=(\partial_{r_+} \Phi)_{\theta,z^i}  \quad , \quad \phi^\theta=(\partial_{\theta} \Phi)_{r_+,z^i}\,.
\end{equation}
It is straightforward to see that the zero point of the vector field i.e., $\phi^{a}(z^i) = 0$, corresponds to the critical points in the thermodynamic phase diagram in $(r_+, T)$ space: specifically, the zeros of the first component $\phi^{r_+}=0$ is equivalent to the critical point
\begin{align}\label{crtical_con}
		\left( \frac{\partial T}{\partial r_+}\right)_{P,z^i} = 0 \quad \text{and} \quad   \left(\frac{\partial^2 T}{\partial r^2_+}\right)_{P,z^i} = 0\,. 
	\end{align}
The zeros of the second component, $\phi^\theta =0$ are always located at $\theta=\pi/2$, which provides a pictorial convenience for the analysis of winding number of the vector field.

\noindent

The proposal in \cite{Wei:2022dzw} suggested that the critical points could be classified into two types based on their topological charge: conventional ($Q_t = -1$) and novel ($Q_t = +1$). According to this proposal, conventional critical points indicate the presence of first-order phase transitions, whereas novel critical points do not.
However, a subsequent study in \cite{Yerra:2022alz} showed that this classification is not universally applicable. Notably, the authors in \cite{Yerra:2022alz} presented a counterexample where a conventional critical point ($Q_t = -1$) failed to exhibit a first-order phase transition, contradicting the original proposal. Furthermore, they employed winding numbers to classify critical points as follows :
\begin{itemize}
	\item $Q_t = -1$: Conventional critical points are associated with phase annihilation point as pressure increases. 
	\item $Q_t = +1$: Novel critical points are associated with phase creation point as pressure increases. 
\end{itemize}

\noindent
\textit{Black hole solutions as defects:} 
This approach employs Duan's potential as a generalized/off-shell free energy 
\cite{York:1986it}.  In the grand canonical ensemble, the generalized Gibbs free energy $\mathcal{G}$ is given through the form of Legendre transformation from the enthalphy $M$ as
\begin{equation}
\label{Generalized_Free_Energy}
\mathcal{G}=M-\frac{S}{\tau} - \varphi Q = \mathcal{G} (r_+,\tau,P,\varphi)\,,
\end{equation}
where $S$ is the entropy, $Q$ is the charge and $\varphi$ is the electric potential. The parameter $\tau$ represents the inverse of the ensemble temperature, which is \textit{a priori} independent of the Hawking temperature $T$, rending $\cal G$ an off-shell quantity. When imposing the on-shell condition given by the extremal condition, the ensemble temperature coincides with the Hawking temperature $\tau^{-1} = T$:
\begin{equation}\label{extremal_free_energy}
    \left(\frac{\partial M}{\partial S}\right)_{P,Q,\alpha} \equiv T = \tau^{-1} \,.
    \end{equation}
We note in \eqref{Generalized_Free_Energy} that the electric potential $\varphi$ is taken to be its on-shell value per \cite{Cai:2001dz, Torii:2005nh}, differing from the approach \cite{Li:2023ppc}. \\

The vector field $\phi$ is then defined  as \cite{Wei:2022dzw}: 
\begin{equation}
\label{Eq:Topological_Defect_Vector_Field}
\phi=\Big( \frac{\partial\mathcal{G}}{\partial r_+}, -\cot\Theta \csc\Theta \Big)\,.
\end{equation}
Here, the auxiliary coordinate $\Theta$ is introduced again to construct a two-dimensional vector field. The zero point of the second component of $\phi$ always corresponds to $\Theta=\pi/2$. Importantly,  the zero points of the first component of  $\phi$ correspond to the on-shell black hole solutions, as they satisfy the extremal condition of the off-shell free energy. 

These defects, corresponding to black hole solutions, are distinguished by their topological charges, given by \eqref{topological_charge}. These charges are determined by the winding numbers of the vectors surrounding the defect locations.
\begin{itemize}
    \item Local stability of black holes: The winding numbers reflect the local stability of black hole branches, with stable branches corresponding to $w_i = +1$ and unstable branches to $w_i =-1$. 
\item Topological classification of black holes: 
The topological charge $W$, obtained as the sum of winding numbers for all branches,  
enables the classification of black hole systems into distinct topological classes: $W = -1\,,0\,, +1$ ~\footnote{We denote the charge as $W$ instead of $Q_t$, to distinguish from the topological charge for the critical point. This classification was recently further refined in  \cite{Wu:2024asq} into four classes $W^{1-}\,,W^{0+}\,,W^{0-}\,,W^{1+}$. However, this paper focuses only on $W =-1,0,+1$.  }. These classes provide universal insight into the patterns of black hole phases.
\end{itemize}
 The local stability implication is evident, as the winding number of the vector field \eqref{Eq:Topological_Defect_Vector_Field} relates the sign of $\left(\frac{\partial^2 \mathcal{G}}{\partial r_+^2}\right)|_{z_i}$ at each zero point (see Appendix in \cite{Liu:2022aqt} for details). 
 We will explicitly investigate the sign of the specific heat in our system and confirm this interpretation. \footnote{In \cite{Li:2023ppc, Li:2023men}, the topology of RN-GB AdS black holes was explored using the black hole radius and charge as order parameters, rather than ($r_+-\Theta$).  While this approach aligned the topological defect of the gradient field with black holes, as proposed in  \cite{Wei:2022dzw}, the stability of these black holes is not necessarily determined by the winding number of the zero points. Specifically, both stable minima and unstable maxima were associated with the same winding number, $+1$, contrary to the expectation that positive and negative winding numbers denote stable and unstable black holes, respectively. 
Interestingly, topological analysis for this order parameters, reveals that source and sink/basin points share the same winding number, $+1$, while saddle points exhibit a winding number of $-1$. Thermodynamic stability was thus determined by the presence of these points, with sink/basin points being thermodynamically stable, whereas source and saddle points are unstable.}
 In addition to the local stability, we will analyze the global stability of our black hole systems to determine the phase structure, as topology alone does not provide a complete picture. For example, the Hawking-Page phase transition is identified by the global minima of the free energy. For further studies on this topic, see \cite{Cho:2002hq, Cai:2007wz, Myung:2008af, Eune:2013qs, Khimphun:2016gsn, Su:2019gby, Kumar:2022fyq, Eom:2022nwc, Yerra:2022coh}.

\section{Charged AdS$_5$ black holes in Einstein-Gauss-Bonnet gravity}\label{review}
The Einstein-Maxwell Gauss-Bonnet action in five
dimensions is given by \cite{Cai:2001dz, Torii:2005nh}
\begin{align}
    S = \frac{1}{16 \pi } \int d^5x \sqrt{-g} \left(R - 2 \Lambda + \frac{\alpha}{2}(R_{\mu\nu\rho\sigma}R^{\mu\nu\rho\sigma}\right.
    \\
\left.    -4R_{\mu\nu}R^{\mu\nu}+R^2)
  -4\pi  F_{\mu\nu}F^{\mu\nu}\right) +S_b \,,\nonumber
\end{align}
where $\Lambda$ is the cosmological constant expressed in terms of AdS radius $\ell$, as $\Lambda = -6 \ell^{-2}$, $\alpha$ is the Gauss-Bonnet coefficient with dimension $(length)^2$  and $S_b$ is the boundary term~\cite{Gibbons:1976ue, Hawking:1995ap, Davis:2002gn, Brihaye:2008xu}. In this work, we restrict ourselves to the cases where $\alpha \geq 0$.
The most general known black hole solution to this theory is given by the charged Gauss-Bonnet asymptotically AdS$_5$ black holes as 
\begin{align}
    ds^2 = -f(r) dt^2 + f^{-1}(r) dr^2 + r^2 d \Omega_3^2 \,,
\end{align}
where the $ d\Omega_3^2$ components of the line element denote a 
$3d$ hypersurface associated with the constants  $k= -1, 0, +1$ corresponding to the curvatures of hyperbolic, planar, and spherical geometry of the black hole horizon, respectively. The corresponding metric function $f(r)$ is given by 
\begin{align}\label{horizon}
   f(r)= k\! +\! \frac{r^2}{2\alpha}\! \left(\!1\!-\! \sqrt{1+\frac{32 \alpha M }{3 \pi r^4} - \frac{16 \alpha Q^2 }{3 \pi^2 r^6}-\frac{4 \alpha}{l^2}}\right)\,.
\end{align}

Recent research has focused on extended thermodynamic space~\cite{Kastor:2009wy,Dolan:2010ha, Dolan:2011xt,Kubiznak:2012wp,Wei:2014hba} for determining criticality in AdS black holes, where the understanding that the negative cosmological constant causes positive vacuum pressure in spacetime entails employing the cosmological constant $\Lambda$ as the thermodynamic pressure~$P$~\cite{Dolan:2010ha, Dolan:2011xt,Kubiznak:2012wp,Wei:2014hba}. We refer \cite{Karch:2015rpa,Mancilla:2024spp} for
the interpretation in the context of AdS/CFT correspondence.
    \begin{align}
        P = -\frac{\Lambda}{8 \pi } = \frac{3}{4 \pi \ell^2} \,.
    \end{align}

    \noindent
 Thus, we present the list of thermodynamic quantities in terms of pressure $P$,  black hole charge $Q$, and horizon radius $r_+$, which is given by the largest positive root of the metric function $f(r)|_{r=r_+}=0$\,
\begin{align} 
 \label{M}  M &= \frac{ \pi }{8}\,\left( 3 k  r_+^2  + 4 \pi P r_+^4 + 3 \alpha k^2 \right) + \frac{ Q^2 }{2 \pi r_+^2} \,,\\
  \label{T} T &= \frac{ \pi^2 r_+^4 (8 \pi P r_+^2 + 3k )- 4 Q^2}{6 \pi^3  r_+^3 (r_+^2 + 2 k \alpha)}\,,
  \\
 \label{S}  S &= \frac{\pi^2 r_+^3}{2 }\left(1+\frac{6 k \alpha}{r_+^2} \right)\,,\\
\label{phi}   \varphi &= \frac{ Q}{\pi r_+^2} \,,\qquad { V= \frac{\pi^2}{2} r_+^4\,, \qquad {\cal A} =\frac{3 \pi^2}{8} k^2\,,}
\end{align}
where electric potential $\varphi$, volume $V$ and ${\cal A}$ denote the conjugate of electric charge~$Q$, pressure $P$ and Gauss-Bonnet coupling $\alpha$ \footnote{ 
In the extended phase space, the thermodynamic quantities satisfy the first law of the black hole thermodynamic as follows \begin{align*}
    d M = T dS + \varphi d Q &+ V d P +\mathcal{A} d \alpha\, \\
    T = \left(\frac{\partial M}{\partial S}\right)_{P,Q,\alpha},&\;V = \left(\frac{\partial M}{\partial P}\right)_{S,Q,\alpha},\;\mathcal{A} = \left(\frac{\partial M}{\partial \alpha}\right)_{S,Q,P},
\end{align*} where $V$ is the thermodynamic volume, the conjugate variable of the pressure $P$ and ${\cal A}$ denote the conjugate Gauss-Bonnet parameter $\alpha$ and while the mass obtains a new physical meaning, the enthalpy $M \equiv H$ rather than the energy of the system.}. 
The black hole entropy deviates from the standard area formula in the presence of higher-derivative corrections, such as those introduced by a non-zero GB coupling $\alpha$, so we follow the formula suggested in \cite{Cai:2001dz}, which is simply given by integrating the first law of thermodynamics under the assumption that entropy vanishes when horizon shrinks to zero.

To deal with grand canonical ensemble, we will use the first relation of the \eqref{phi} to express the system in terms of the electric potential, such that
\begin{align} 
 \label{Mgrand} M &= \frac{ \pi }{8}\,\left( (3 k  + 4 \varphi^2)r_+^2  + 4 \pi P r_+^4 + 3 \alpha k^2 \right) 
   \,,
   \\ 
   \label{Tgrand}
  T &= \frac{8 \pi  P r_+^3 + \left(3 k-4  \varphi^2\right) r_+}{6 \pi(r_+^2 + 2k  \alpha) }\,.
\end{align}

We note from the above expression that the Gauss-Bonnet parameter $\alpha$ is always accompanied by constant curvature $k$ thus for $k=0$  thermodynamic quantities do not get affected by the presence of the Gauss-Bonnet term. 

We should note that the parameters must satisfy certain limits for well-defined solutions. 
For a given $\alpha>0$, the first constraint comes from the metric function \eqref{horizon}, where a well-behaved asymptotic vacuum solution, which is realized by $M=Q=0$,  requires
\begin{align} \label{Pmax}
  0 \leq P \leq  \frac{3}{16 \pi \alpha }\equiv P_{\max}\,.
\end{align}
Equivalently, for given pressure $P$,  the Gauss-Bonnet coupling constant is bounded from above, $\alpha \leq 3/(16\pi P) \equiv\alpha_{\max}$. 

The range of the black hole radius $r_+$ is also constrained by the following conditions: Firstly, the physical condition imposed by the non-negative temperature \eqref{Tgrand} restricts the value of the horizon radius,
\begin{equation} \label{cons_temp}
    r_+ \geq \sqrt{\frac{4\varphi^2 -3k}{8\pi P}} \equiv r_{T=0}\,.
\end{equation}
Secondly, the non-negative definiteness of the black hole entropy 
\eqref{S} \cite{Cvetic:2001bk, Nojiri:2002qn} demands \footnote{ We do not consider the possibility of adding an ambiguity to redefine the entropy \cite{Cvetic:2001bk,Clunan:2004tb}.}
\begin{align}\label{enon}
r_+^2 \geq - 6k \alpha \equiv r_{S=0}^2\,.
\end{align}
From \eqref{cons_temp} and \eqref{enon}, we infer that the spherical curvatures $(k= +1)$ allow for all the positive values of horizon radius as far as $\varphi^2 \leq {3}/4$. In the case where $\varphi^2 > 3/4$, or for the planner and hyperbolic horizon $(k = 0, -1)$, the radius is bounded from below depending on the value of~$\varphi$.   
\footnote{This result will be interchanged between spherical and hyperbolic geometry for $\alpha < 0$.
There are studies on the solutions and properties with both signs of $\alpha$~\cite{Koh:2014bka, Lee:2018zym, Lee:2021uis, Biswas:2023eju, Biswas:2024viz}.
} Keeping the constraints in mind we will study the topological thermodynamics for all three geometries in the following \autoref{topology}.

\section{Topological natures of the black hole in grand canonical ensemble}
\label{topology}
In this section, we investigate the nature of 5-dimensional RN-GB AdS black holes in the grand canonical ensemble using Duan's topological $\phi$-mapping theory, as reviewed in Section \autoref{PhimapTheory}, and explore its implications through two distinct approaches. In \autoref{Critical_Point_Topology}, we follow the analysis in \cite{Wei:2021vdx} and categorize the  nature of critical points based on \cite{Yerra:2022alz}.  
In \autoref{Black_Holes_Defects}, we follow \cite{Wei:2022dzw}, and consider the generalized free energy  to study the local stability of black hole phase based on the sign of the corresponding winding number.  In the end,  the topological class of the RN-GB AdS black hole in the grand canonical ensemble is investigated based on their topological number.


\subsection{ Topology of critical point }\label{Critical_Point_Topology}
In this approach, one first calculates the extremal points of the temperature \eqref{Tgrand}. As explained in the~\eqref{duan_poten}, we use the condition $\left(\frac{\partial T}{\partial r_+}\right)_{P,\varphi}=0$ and  yield an expression for pressure given as,
    \begin{align}
        P= \frac{\left(3 k-4  \varphi^2\right) \left(r_+^2-2k \alpha  \right)}{8 \pi  r_+^2 \left(r_+^2 + 6 k \alpha  \right)}\,.
    \end{align}

\noindent    
Plugging this into \eqref{Tgrand}, we obtain the temperature $T (r_+,z^i)$ of these extremal points and define the  Duan's potential as 
\begin{align}
    \Phi = \frac{r_+ \csc \theta  \left(3 k-4  \varphi^2\right)}{3 \pi  \left(r_+^2 + 6 k \alpha  \right)}\,.
\end{align}

The corresponding vector field $\phi = (\phi^{r_+},\phi^\theta)$ is obtained as
\begin{align}\label{vector_gc1}
    \phi^{r_+} &= \left( \frac{\partial \Phi}{\partial r_+}\right)_{\varphi, \theta}= -\frac{(r_+^2 - 6k\alpha)   \left(3 k-4 \varphi^2\right) \csc \theta}{3 \pi  \left(r_+^2 + 6 k \alpha  \right)}\,,\\
    \label{vector_gc2} \phi^{\theta} &= \left( \frac{\partial \Phi}{\partial \theta}\right)_{\varphi, r_+}= -\frac{r_+  \left(3 k-4  \varphi^2\right) \cot \theta \csc \theta}{3 \pi  \left(r_+^2 + 6 k \alpha  \right)}\,,
\end{align}
and the normalized vector components are given by the unit vector $n^a = \frac{\phi^a}{||\phi^a||} $ as
\begin{align}
  n^{r_+} &
  =  -\frac{r_+^2-6 k\alpha  }{\sqrt{r_+^2 \cot ^2 \theta \left(r_+^2 + 6k \alpha  \right)^2+\left(r_+^2-6 k \alpha  \right)^2}}\,,\\
   n^{\theta} &
   =  -\frac{r_+(r_+^2+6 k\alpha) \cot \theta }{\sqrt{r_+^2 \cot ^2 \theta \left(r_+^2 + 6k \alpha  \right)^2+\left(r_+^2-6 k \alpha  \right)^2}}\,.
\end{align}

\noindent

In this construction, the critical points in the phase diagram defined in \eqref{crtical_con} are determined by setting the \eqref{vector_gc1} and  \eqref{vector_gc2} to zero. We find one critical point at 
\be \label{CritPt}
(r_c\,, \theta )=(\sqrt{6 k \alpha},\pi/2)\,,
\ee
and thus the critical pressure and the temperature are given as 
\begin{align}\label{critical_P_T}
    P_c = \frac{3 k-4  \varphi^2}{144 \pi  \alpha  k} \quad \text{and} \quad T_c = \frac{3 k-4 \varphi ^2}{6 \pi \sqrt{6 k \alpha} } \,.
\end{align}
Based on the above analysis, we observe that:
\begin{itemize}
\item 
The RN-AdS black holes without Gauss-Bonnet coupling do not exhibit criticality in the grand canonical ensemble which can be checked by taking $\alpha$ going to zero limit\footnote{In the canonical ensemble, however, there exists a critical point in this limit given as
\begin{align*}
    r_c= \frac{\sqrt{2} \sqrt[4]{5} \sqrt{Q}}{\sqrt{\pi } \sqrt[4]{k}}.
\end{align*}}. This agrees with the study in~\cite{Zou:2014mha}. 
\item For $k=0$ case, no critical points were found.  It is also evident that there cannot be a physical solution for $r_c$ for $k = -1$. This means that the black hole horizon with planar or hyperbolic geometry does not have a critical point.
    \item For $k=1$ and $\alpha>0$, there is a critical point~$r_c$, which is independent of the choice of electric potential. However, critical temperature $T_c$ and critical pressure $P_c$ depend on the potential such that, for the potential $\varphi > \sqrt{3}/{2}$, they both disappear or become negative. This concludes that, in a grand canonical ensemble, the RN-GB AdS$_5$ black holes with spherical horizon show critical behavior in the potential range $ 0 \leq \varphi\leq \varphi_c^{\text{max}}\,,$ where $ \varphi_c^{
    \text{max} }\equiv {\sqrt{3}}/{2}$.
\end{itemize}

\noindent
\textit{Nature of critical point:} Depending on the sign of the topological charge $Q_t$, critical points can be categorized into two types, as discussed in \cite{Wei:2021vdx, Yerra:2022alz}. Specifically, the novel critical point corresponds to $Q_t=1$, while the conventional critical point corresponds to $Q_t=-1$, a scenario that applies to our system.

To determine the topological charge of a critical point (where $\phi^a = 0$), we compute its winding number $(w_i)$  defined in 
\eqref{topological_charge} as
 \begin{align}
     Q_t = \frac{1}{2 \pi} \Omega(2\pi)\,,
 \end{align}
 where $\Omega(\vartheta)$ of the vector field $\phi$ along a given contour as
\begin{equation}
\label{Deflection Angle}
\Omega(\vartheta)\equiv \int_0^\vartheta \epsilon_{ab}n^a\partial_\vartheta n^b  d\vartheta\,.
\end{equation}

\begin{figure}[ht]
	\centering
	\includegraphics[width=.35\textwidth]{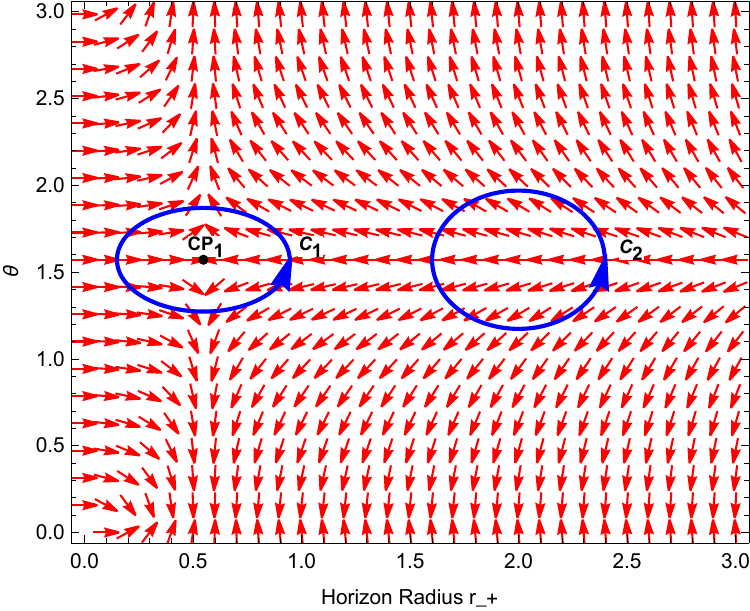}
	\caption{\footnotesize Plot of the normalized vector field $n^a$ in $r_+$ vs $\theta$ plane for charged Gauss-Bonnet AdS black hole in the grand canonical ensemble. The contour $C_1$ encloses the critical point denoted as $CP_1$, and the $C_2$ does not.}
	\label{Fig:Critical_Point}
	\end{figure}

To compute, we choose a contour $C$ on the $r_+ -\,\theta$ plane to be an ellipse centered at $(r_0, \frac{\pi}{2})$ parameterized by the angle $\vartheta\in(0,2\pi)$ 
\begin{equation}	
\label{Contour}
	\begin{cases}
		&r_+=c \cos\vartheta+r_0 \,,\\
		&\theta=d \sin\vartheta+\frac{\pi}{2}\,,
	\end{cases}
\end{equation}
where $c\,,d$ are constants for the shape of the ellipse.
In \autoref{Fig:Critical_Point}, we show two test contours $C_1$ and $C_2$, where the $C_1$ encloses the critical point $CP_1$ given by \eqref{CritPt} whereas $C_2$  is outside of the critical point. 
In \autoref{Fig:Critical_Point}, we show the critical point for a choice $\alpha=0.05 $, and  for computing the topological charge for this critical point, we choose $(c,d,r_0)=(0.3\,,0.2\,,\sqrt{0.3}= r_c)$ and $(0.4\,,0.4\,,2 \neq r_c)$ for contours $C_1$ and $C_2$, respectively.
\begin{table}[ht]
	\centering{	
		\begin{tabular}{|c|c|c|c|c|c|c|}
			\hline \hline 
			\multicolumn{2}{|c|}{Case} &  $C_1$ & $C_2$ & $Q_1$ & $Q_2$ & $Q$ \\ \hline  
			\multirow{3}{4em} {$\alpha = 0$} & c & $0.07$ & $0.07$ & \multirow{3}{2em} {$0$} & \multirow{3}{2em} {0}& \multirow{3}{2em} {$0$} \\ 
			& d & $0.4$ & $0.4$ & &  &  \\ 
			& $r_0$ & $0.5\neq r_c$ & $1\neq r_c$ &  &  &  \\ \hline
			\multirow{3}{4em} {$\alpha = 0.05$} & c & 0.3 & 0.4 & \multirow{3}{2em} {$-1$}&\multirow{3}{2em} {0} & \multirow{3}{2em} {$-1$} \\ 
			& d & 0.2 & 0.4 &  &  &  \\ 
			& $r_0$ & $0.5577 = r_c$ & $2\neq r_c$ &  &  &  \\ \hline
			\multirow{3}{4em} {$\alpha = 0.5$} & c & 0.3 & 0.15 & \multirow{3}{2em} {$-1$} & \multirow{3}{2em} {0} & \multirow{3}{2em} {$-1$} \\ 
			& d & 0.2 & 0.4 &  &  &  \\ 
			& $r_0$ & $1.732= r_c$ & $1\neq r_c$ & &  &  \\ 
			\hline\hline 
		\end{tabular}
		\caption{        
       \footnotesize  Topological charge of critical points for various Gauss-Bonnet coupling in grand canonical ensemble.}
		\label{table:critical_point} }
\end{table}

 \noindent
 
We found that for the critical point $CP_1$ enclosed by the contour $C_1$,  the topological charge is $Q_{CP_1}=-1$ which means it is a conventional critical point. Since the contour $C_2$ does not enclose any critical point, it corresponds to zero topological charge. Thus, the total topological charge is $Q_t=-1$. Thus, unlike RN AdS black hole, RN-GB AdS black exhibits one critical point both in canonical and grand canonical ensemble.
In \autoref{table:critical_point}, we provide the list of critical points and topological charges for different values of Gauss-Bonnet coupling $\alpha$. 
\begin{figure} [ht]
	\centerline{
	\includegraphics[scale=0.4]{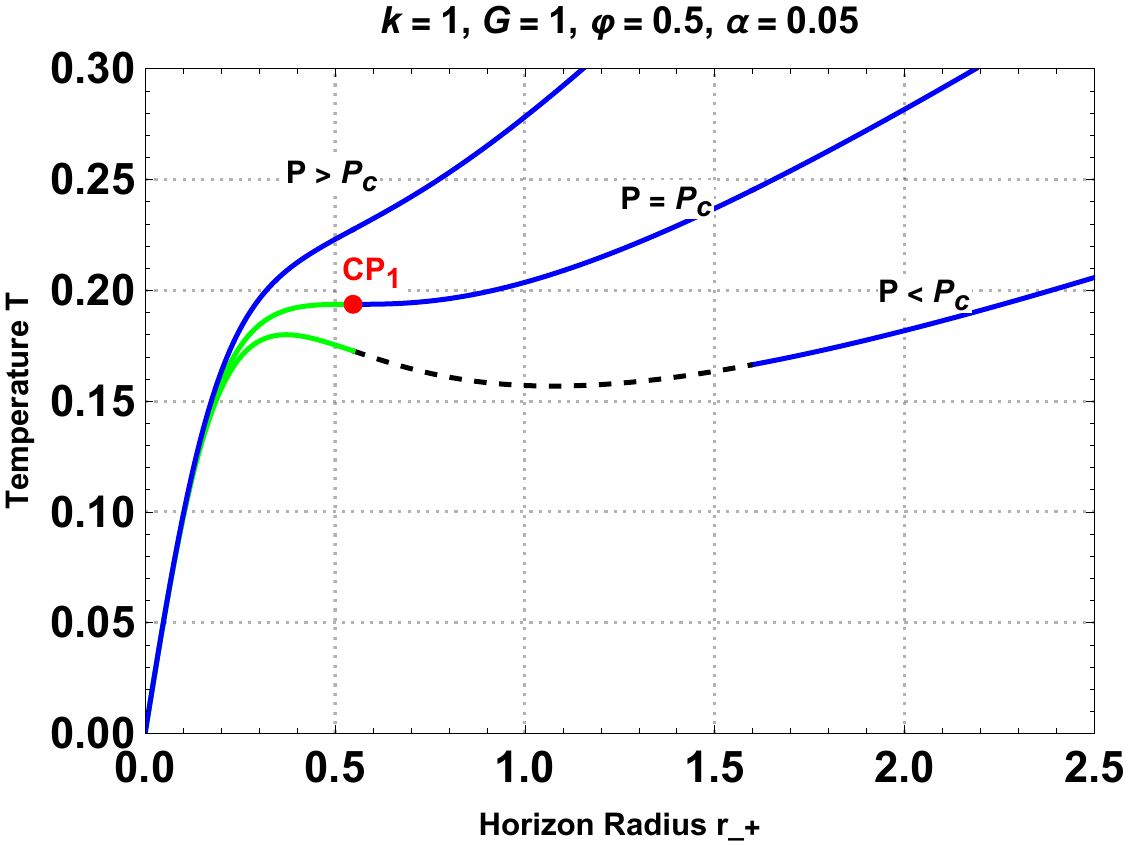}}
	\caption{\footnotesize  Isobaric curves of RN-GB AdS black hole in the grand canonical ensemble, showing the disappearance of phases near the critical point $CP_1$. The dashed curves denote unstable black hole branches and solid curves denote
stable black hole branches.}
	\label{Fig:Isobaric_Curve}
	\end{figure}

\noindent
In Fig\eqref{Fig:Isobaric_Curve}, we plot the $T-r_+$ isobaric diagram for spherical RN-GB AdS black holes with a choice $\varphi=0.5 < \varphi_c^{\max}$ and $\alpha=0.05< \alpha_{\max}$. It shows that we have three black hole phases, small stable (green), unstable intermediate (dotted), and large stable (blue) when the pressure is lower than the critical value $P<P_c$. On increasing the pressure, the unstable phase disappears near the critical point $CP_1$, and we are left with only one black hole phase, beyond the critical pressure  $P>P_c$. This is the reason the conventional critical points  $(Q_t|_{cp_1}=-1)$ are classified as a phase disappearing/annihilation point. 
    
\subsection{Topology of the black hole as a defect}\label{Black_Holes_Defects}
In this approach, we start with the generalized Gibbs free energy $\mathcal{G}(r_+,\tau, P,\varphi)$ defined in \eqref{Generalized_Free_Energy}. 
For the RN-GB AdS black hole, it is given as
\begin{eqnarray}
\mathcal{G} &=& \frac{\pi^2}{2} P r_+^4 +\frac{3\pi k}{8} (r_+^2 + k \alpha)- \frac{\pi}{2} r_+^2\varphi^2 \\
&&-\frac{\pi^2 r_+ (r_+^2 + 6k\alpha)}{2\tau}\,.\nonumber
\end{eqnarray}
When the ensemble temperature $\tau^{-1}$ equals the Hawking temperature $T$, the on-shell black hole branches represent the extremal points in the Gibbs free energy as explained in \eqref{extremal_free_energy}. Following \eqref{Eq:Topological_Defect_Vector_Field}, the vector component yields
\begin{eqnarray}\label{vector_defects}
\phi^{r_+}  &=& {2\pi^2} P r_+^3 +\frac{3\pi k}{4} r_+ - {\pi} r_+ \varphi^2 \\
&&-\frac{3\pi^2 (r_+^2 + 2k\alpha)}{2\tau}\,,\nonumber
\\
\phi^{\Theta} &=& - \cot \Theta \csc \Theta, \label{vector_phitheta}
\end{eqnarray}
and corresponding unit vectors can be obtained using $n^a = \frac{\phi^a}{||\phi^a||} \,,$ $ a = (r_+, \Theta)$. 
The $\Theta$ direction of the vector $\phi^\Theta$ is zero trivially at $\Theta=\pi/2$. At the zero point of the vector, $\phi^{r_+}=0$, the expression for $\tau$ (inverse of black hole temperature) is obtained as
\begin{align}\label{tausolution}
  \tau = \frac{6 \pi  \left(r_+^2 + 2 k \alpha  \right)}{r_+ \left(3 k+8 \pi  P r_+^2-4 \varphi^2\right)} \,. 
\end{align}
The solutions of this equations represent the branches of the black hole solutions. 
In the following subsections, we will analyze the topological nature of the zero points and local stability for the corresponding black holes for $k= 0, +1, -1$, i.e. for the case of zero, positive and negative curvature of the black hole horizon,  respectively. 


\subsubsection{$k=0$} \label{Sec:k=0_topology}
We recall that the thermodynamic quantities, \eqref{Mgrand}, \eqref{Tgrand} and \eqref{S},  for $k=0$, are interestingly independent of the Gauss-Bonnet parameter $\alpha$. Likewise, the solution  \eqref{tausolution} for planar horizon given by
\begin{align}
  \tau = \frac{3 \pi r_+ }{2 \left(2 \pi  P r_+^2- \varphi^2\right)} , 
\end{align}
\begin{figure}[ht]
\begin{minipage}{0.6\linewidth}
    \centering
    \includegraphics[width=1\textwidth]{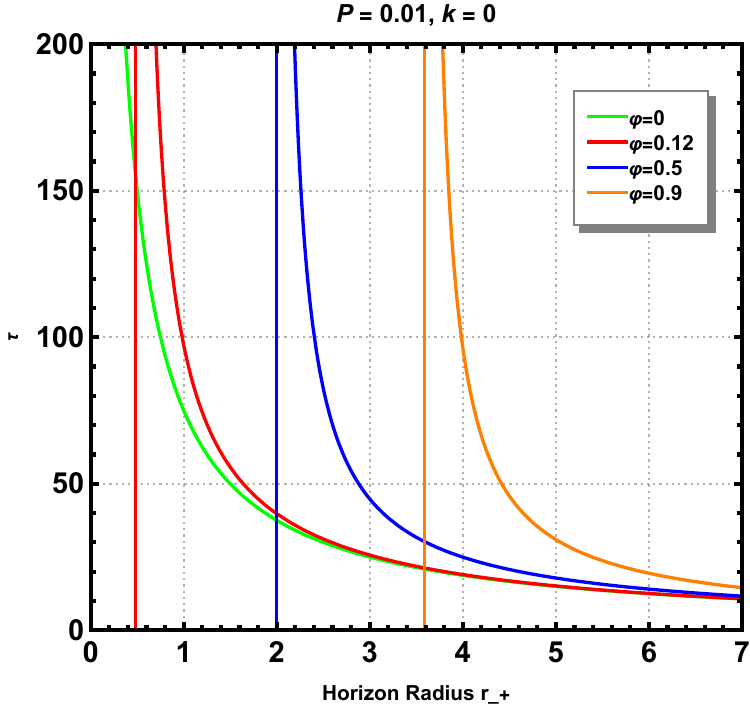}
    \subcaption{}
    \label{k=0_tau_r}
\end{minipage}
\hspace{0.1cm}
\begin{minipage}{0.6\linewidth}
    \centering
   \includegraphics[width=1\textwidth]{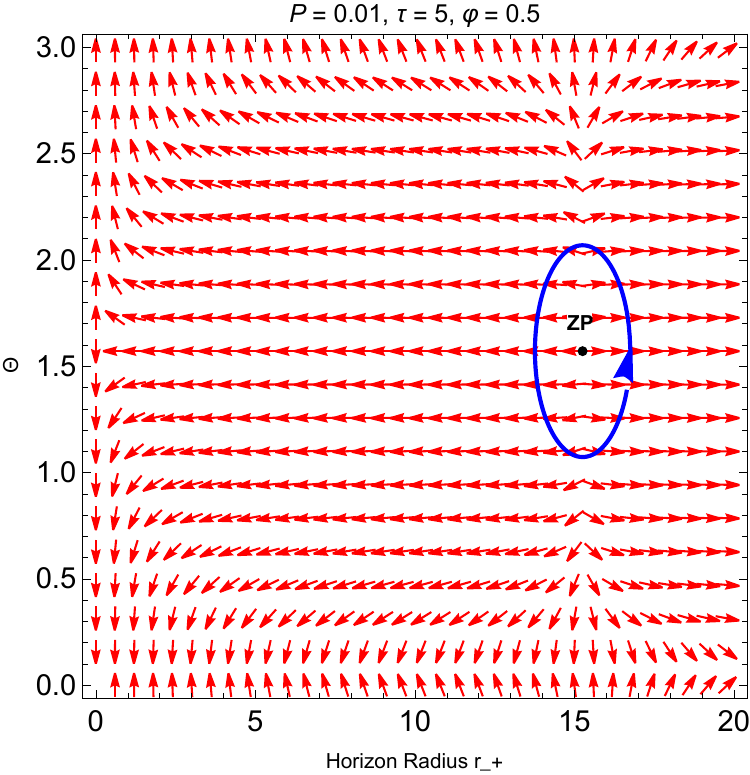}
    \subcaption{}  
      \label{k=0_zero_point}
\end{minipage}
   \caption{\footnotesize The zero points of $\phi^{r_+}$ in (a)  $r_+$- $\tau$ plane and (b)   $r_+$- $\Theta$ plane, for charged AdS black hole in grand canonical ensemble with $\tau=5$, $\varphi=0.5$, and pressure $P=0.01$.} \label{Fig:k=0_topology}
\end{figure}

\noindent
is independent of parameter $\alpha$. In this case, we have only one solution of $r_+$ i.e., one branch of black hole solution. In other words, there is only one zero point of the vector, \eqref{vector_defects} and \eqref{vector_phitheta}. Around the zero point, the winding number of the vector field turns out to be $w= +1$ for any value of the pressure $P$ and  electric potential~$\varphi$. This indicates that for case $k=0$, the black holes are locally thermodynamically stable.   

To illustrate, we show in  \autoref{Fig:k=0_topology}, the branch of the zero points of $\phi^{r_+}$ in  $r_+ - \tau$ plane for various choices of $\varphi$ with a choice of $P=0.01$. The unit vectors in $r_+ - \Theta$ plane and the location of the zero point is also plotted for a choice of $\tau=5$, $\varphi=0.5$, and pressure $P=0.01$. 

\subsubsection{$k=1$}\label{Sec:k=1_topology}
\begin{figure}[ht]
\begin{minipage}{0.38\textwidth}
    \centering
    \includegraphics[width=1\textwidth]{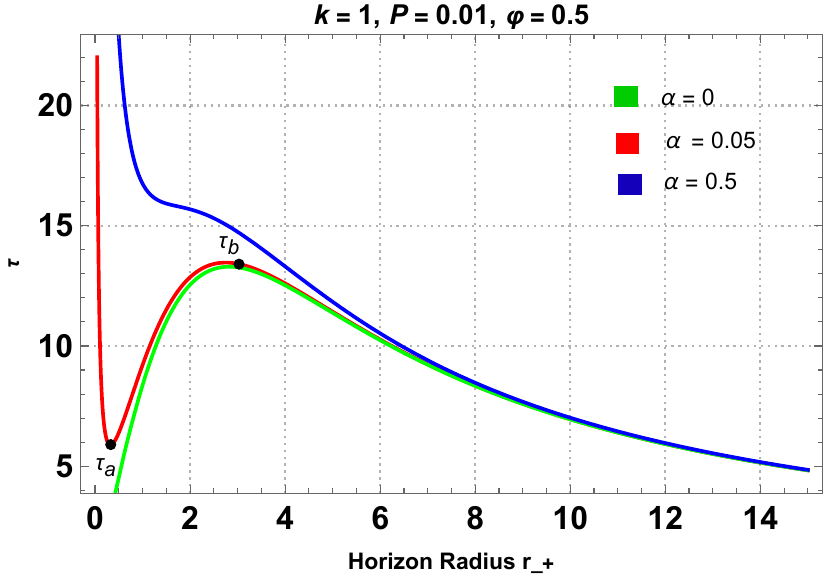}
    \subcaption{}
    \label{k=1_tau_r_low_potential}
\end{minipage}
\hspace{0.1cm}
\begin{minipage}{0.38\textwidth}
    \centering
   \includegraphics[width=1\textwidth]{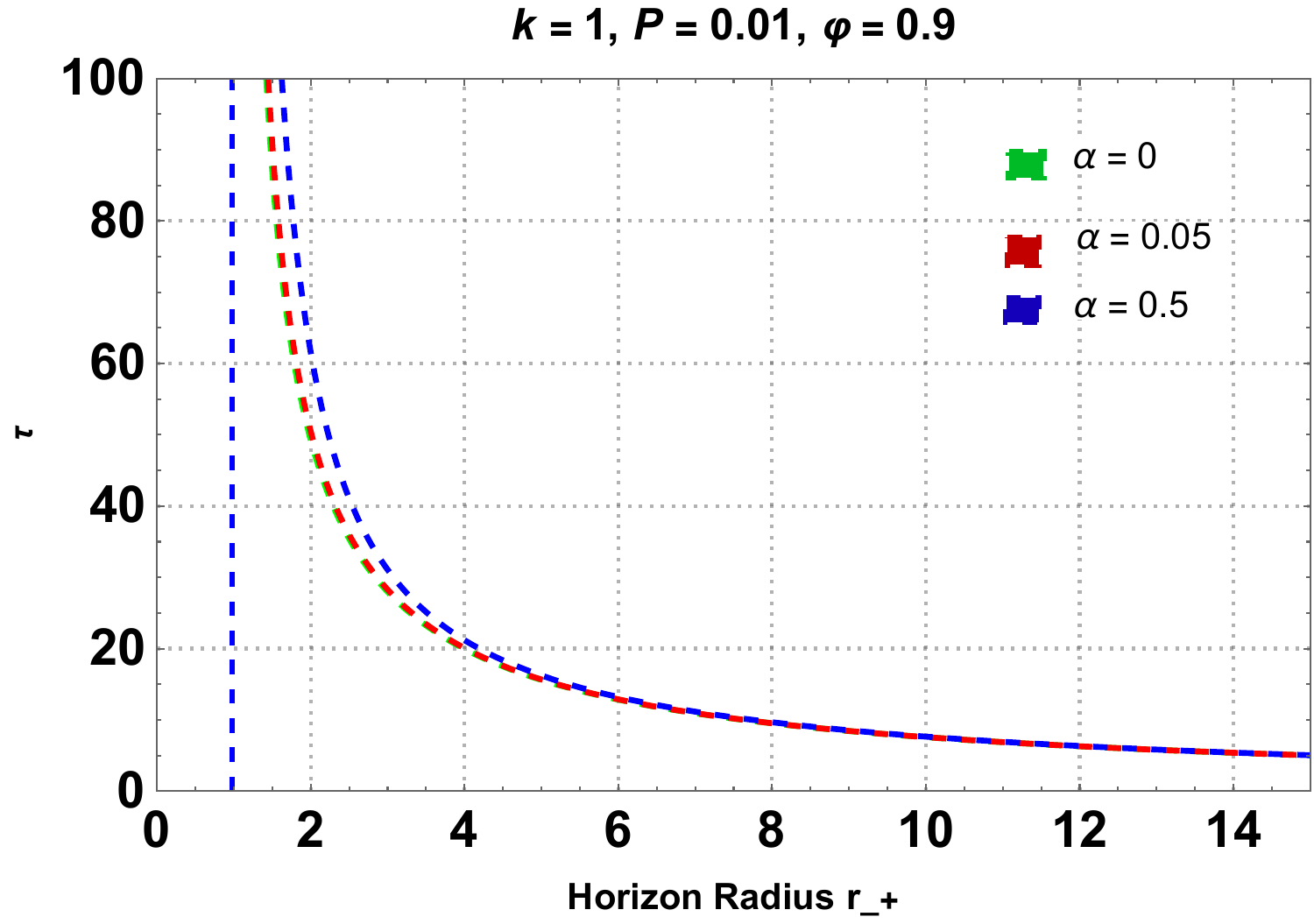}
    \subcaption{}  
      \label{k=1_tau_r_high_potential}
\end{minipage}
   \caption{\footnotesize The zero points of $\phi^{r_+}$ in $\tau$ vs $r_+$ plane for charged AdS black hole in grand canonical ensemble with  $\tau=10$ and pressure $P=0.01$. Here for (a) $\varphi=0.5$, for (b) $\varphi=0.9$  } \label{Fig:k=1_tau_r_topology}
\end{figure}

For $k=1$, the solution~\eqref{tausolution} for the zero point of \eqref{vector_defects} and \eqref{vector_phitheta} becomes 
\begin{align}
  \tau = \frac{6 \pi  \left(r_+^2 + 2  \alpha  \right)}{r_+ \left(3 +8 \pi  P r_+^2-4 \varphi^2\right)} \,. 
\end{align}
The number of solutions for $r_+$ depends on the value of the other parameters.
The \autoref{Fig:k=1_tau_r_topology} plots the branch of solutions in $r_+-\tau-$ plane for various Gauss-Bonnet coupling $\alpha$ and choices of electric potential $\varphi=0.5 < \varphi_c^{\max}$ and $\varphi=0.9> \varphi_c^{\max}$, with choosing $P=0.01$ without loss of generality.

\noindent

For $\varphi =0.5< \varphi_c^{\max}$ (see \autoref{k=1_tau_r_low_potential}), we observe two black hole branches for $\alpha=0$ (RN AdS black holes), three black hole branches for $\alpha =0.05 < \alpha_c$, and only one black hole branch for $\alpha=0.5> \alpha_c$, where the critical value $\alpha_c =0.44$ is determined by the relation~\eqref{critical_P_T}.
For $\varphi = 0.9> \varphi_c^{\max}$ (see \autoref{k=1_tau_r_high_potential}), there is only one black hole branch for all values of $\alpha$. 
\\

For the former case $(\varphi= 0.5)$, the number of zero points of the vector for various $\alpha$ and $\tau$ ranges is summarized in \autoref{table:zero_point_table_spherical}. Below, we discuss each case in detail:

\begin{table}[ht]
	\centering
	\begin{tabular}{|c|c|c|c|}
\hline \hline
\multicolumn{2}{|c|}{Case $k=1$} &  $P= 0.01$ & \# of ZPs  \\ \hline  
\multirow{1}{4em} {$\alpha = 0$} &  & $1-13$ & 2  \\
\& & $\tau$ &  &      \\
{$\varphi=0.5$}& & {$14-\tau_{max} $} & $0$   \\ \hline
\multirow{1}{4em} {$\alpha = 0.05$} &  & $1-5.5$ & $1$   \\
\& & $\tau$ & $5.5-13.5$ & $3$    \\
{$\varphi=0.5$}&  & $13.5-\tau_{max}$ & $1$    \\ \hline
\multirow{1}{4em} {$\alpha = 0.5$} &  &\multirow{3}{4em} {$1-\tau_{max}$} & \multirow{3}{1em} {1}  \\
\& & $\tau$ &  &   \\
{$\varphi=0.5$} &  &  &     \\
\hline \hline
\end{tabular}
\caption{\footnotesize Zero points of spherical (k=1) Gauss-Bonnet charged AdS black holes for various Gauss-Bonnet coupling and the ranges of the inverse temperature $\tau$ in the grand canonical ensemble.} 
\label{table:zero_point_table_spherical}
\end{table}

\begin{figure}[ht]
\begin{minipage}{0.6\linewidth}
    \centering
    \includegraphics[width=1\textwidth]{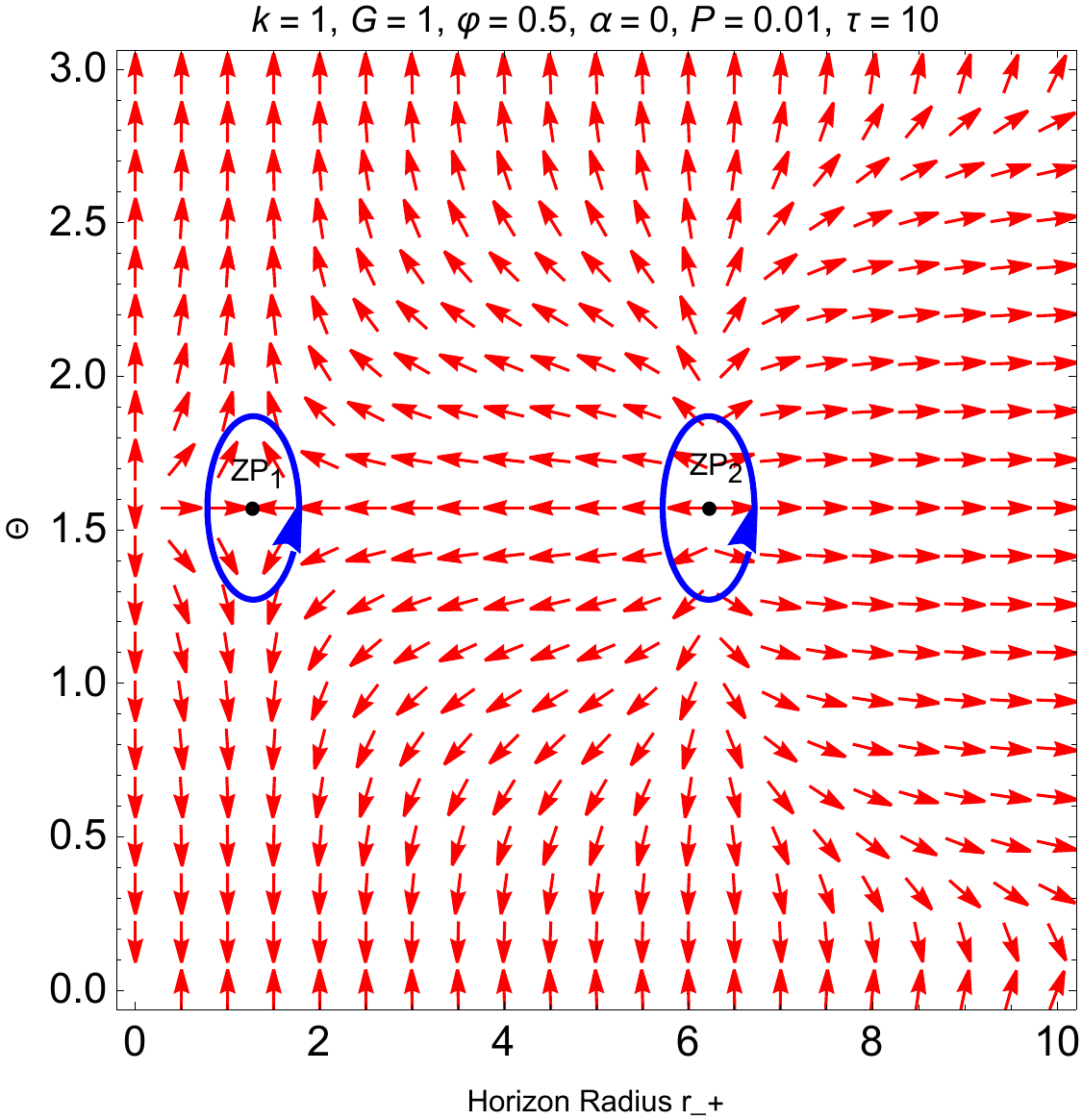}
    \subcaption{}
    \label{k=1_zero_point_low_potential}
\end{minipage}
\hspace{0.1cm}
\begin{minipage}{0.6\linewidth}
    \centering
   \includegraphics[width=1\textwidth]{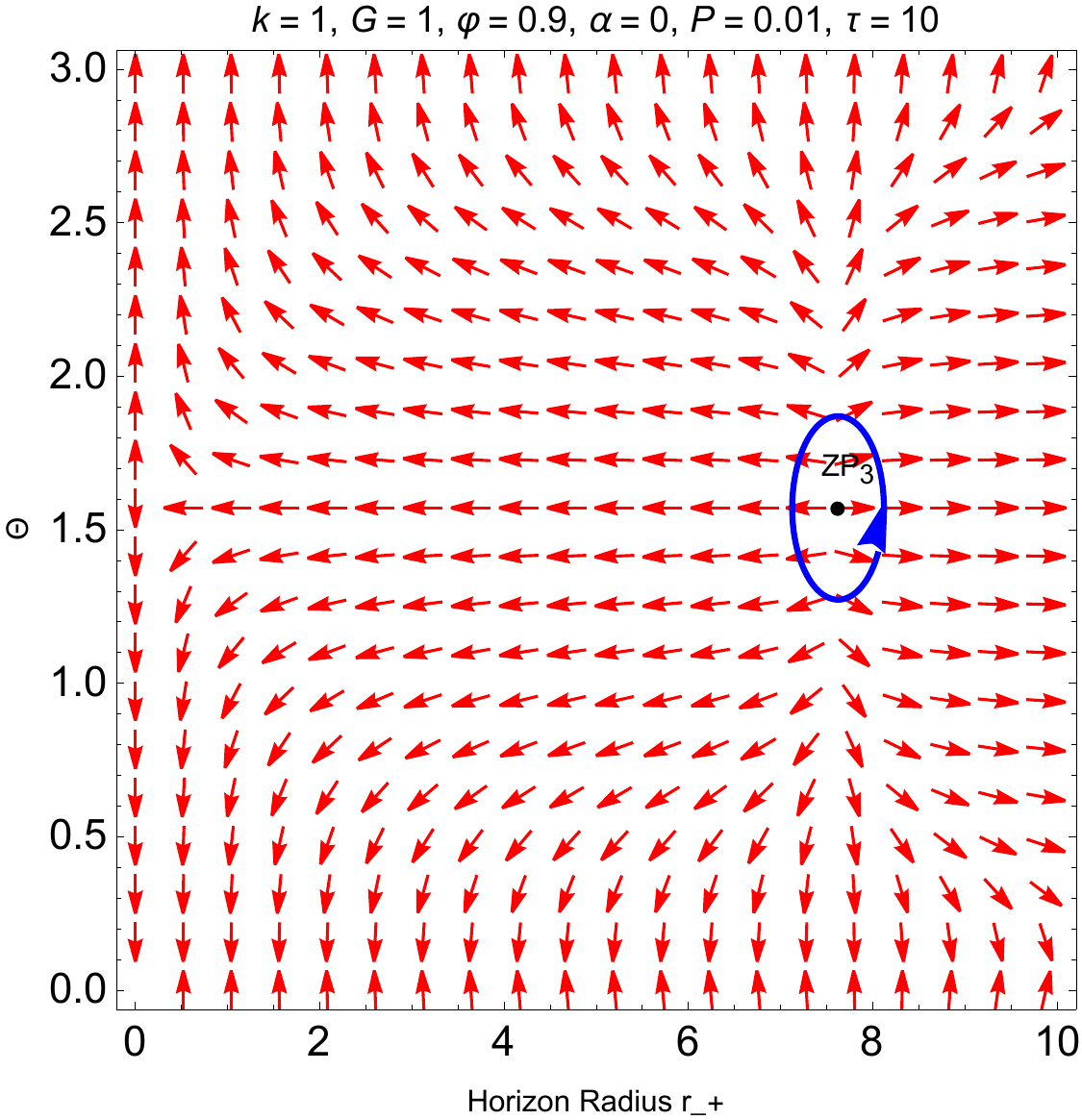}
    \subcaption{}  
      \label{k=1_zero_point_high_potential}
\end{minipage}
   \caption{\footnotesize The unit vectors ($n^1,n^2$) in $\Theta$ vs $r_+$ plane for charged  AdS black hole in grand canonical ensemble with  $\tau=10$ and pressure $P=0.01$. Here for (a) $\varphi=0.5$, for (b) $\varphi=0.9$.} \label{Fig:k=1_zero_point_topology}
\end{figure}

\noindent
{\bf Case 1: $\alpha=0$} \\
As shown by the green curve in \autoref{k=1_tau_r_low_potential}, there is one extremal value of $\tau$ at $\tau_b$. For $\tau=10<\tau_b$, $\varphi=0.5< \varphi_c^{\max}$, and $P=0.01$, two zero points, $ZP_1$ and $ZP_2$, are found, with corresponding winding numbers, $w = -1$ and $ w= 1$, respectively, resulting in a total winding number $W=0$, , which
differs from the case in the canonical ensemble. These zero points are depicted in the $r_+ - \Theta$ plane in \autoref{k=1_zero_point_low_potential}. 
Interestingly, however for potential values $\varphi>0.86= \varphi_c^{\max}$, there is only one black hole branch with a winding number $w=+1$, as shown in \autoref{k=1_zero_point_high_potential}. This gives a total topological number $W = 1$, consistent with the canonical ensemble. 

\begin{figure}[h]
\begin{minipage}{0.6\linewidth}
    \centering
    \includegraphics[width=1\textwidth]{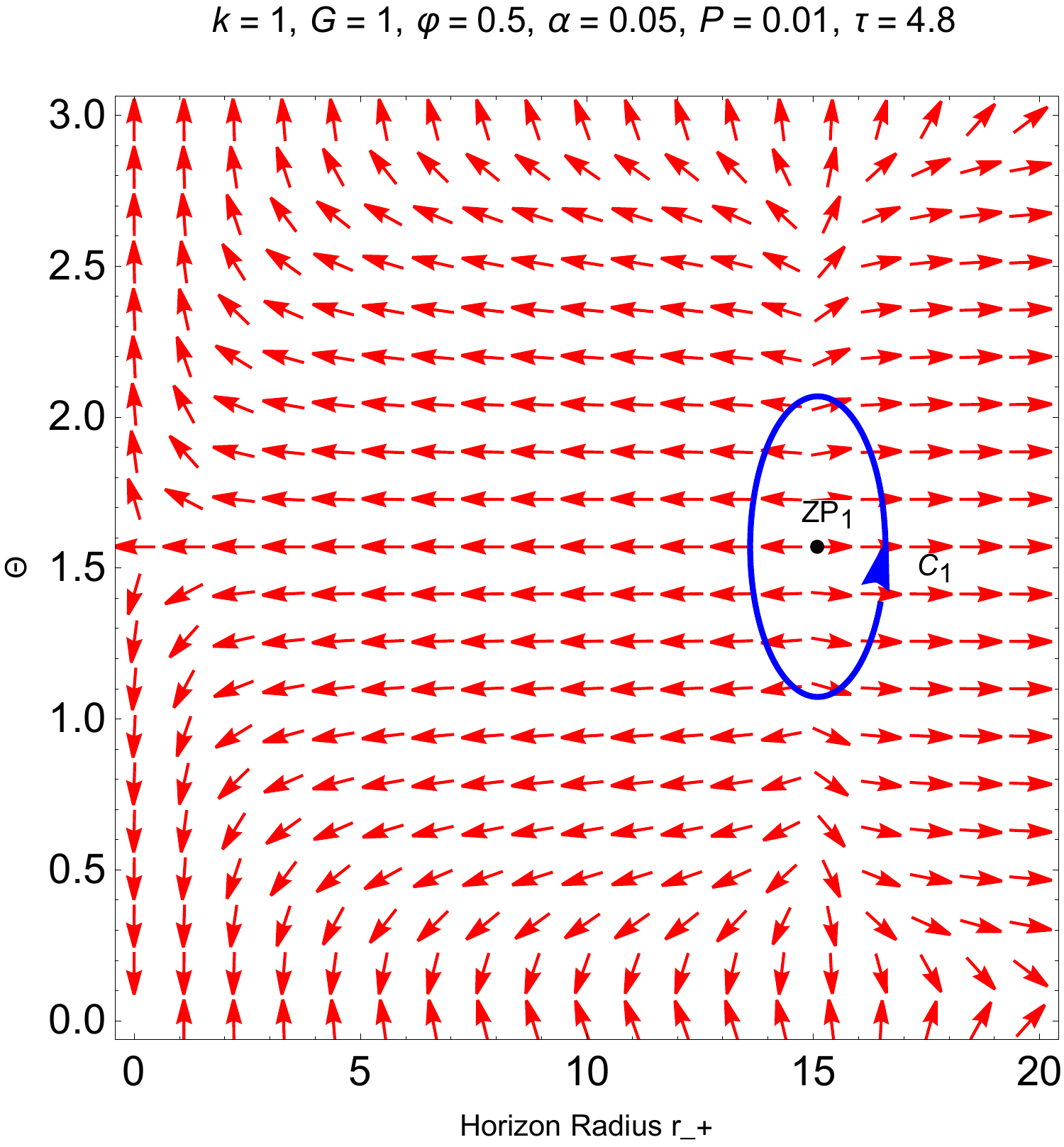}
    \subcaption{}
    \label{Plot_a}
\end{minipage}
\hspace{0.1cm}
\begin{minipage}{0.6\linewidth}
    \centering
   \includegraphics[width=1\textwidth]{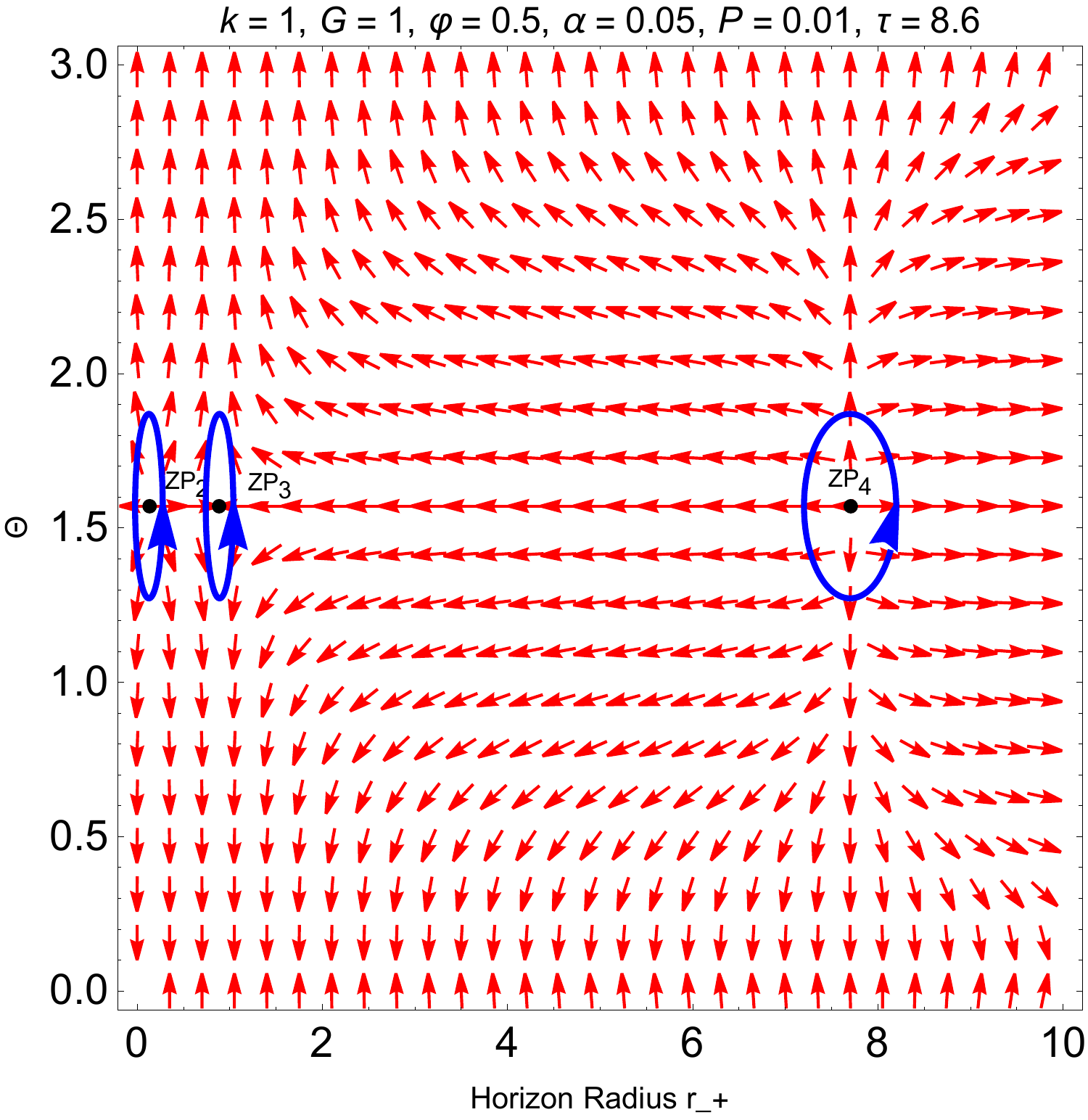}
    \subcaption{}  
      \label{Plot_b}
\end{minipage}
\hspace{0.1cm}
\begin{minipage}{0.6\linewidth} 
    \centering
    \includegraphics[width=1\textwidth]{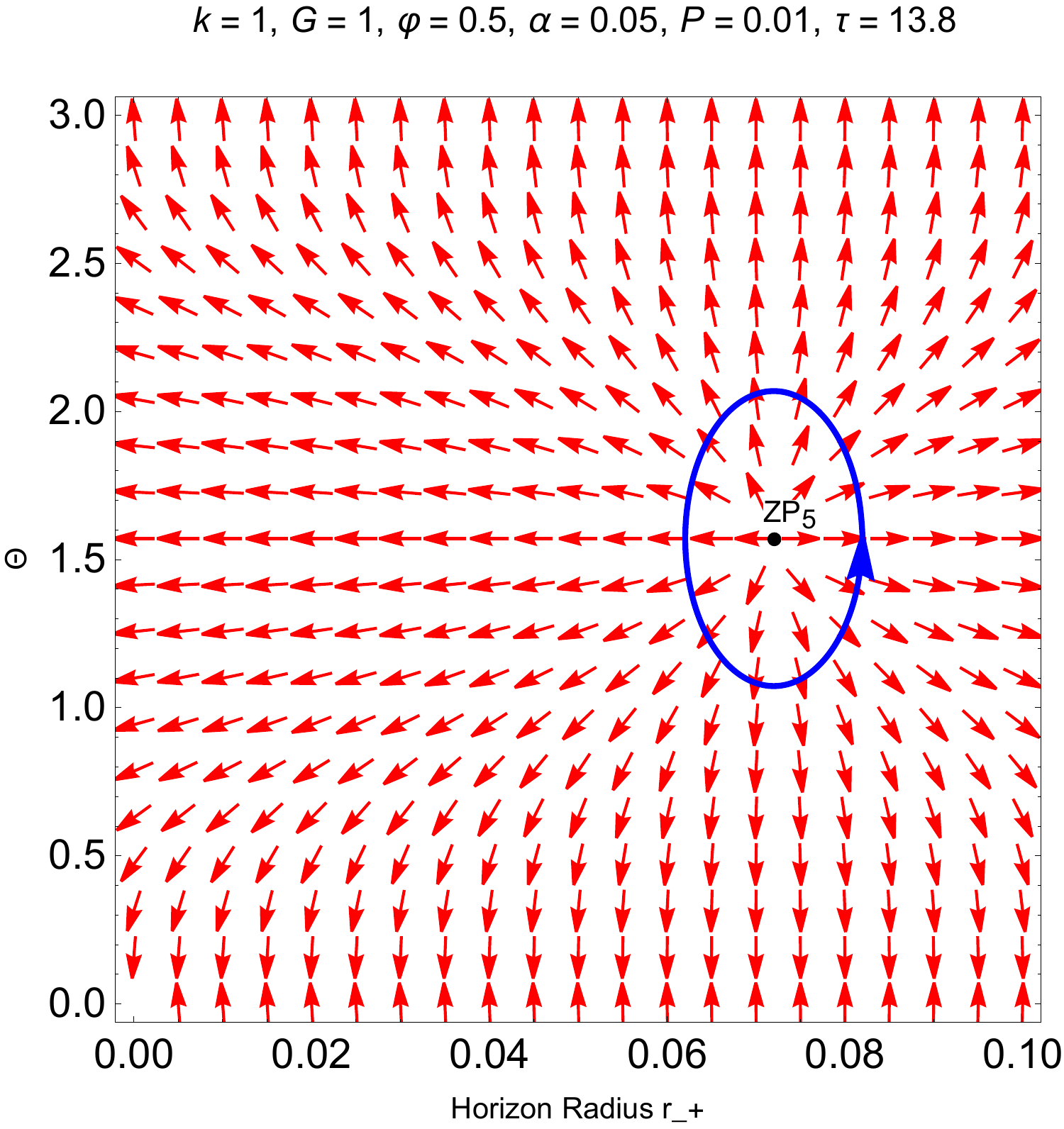}
    \subcaption{}
      \label{Plot_c}
\end{minipage}  
   \caption{\footnotesize The unit vectors ($n^1,n^2$) in $\Theta$ vs $r_+$ plane for charged Gauss-Bonnet AdS black hole in grand canonical ensemble with coupling $\alpha=0.05$ and pressure less than the critical value. Here for (a) $\tau=4.8$, for (b) $\tau=8.6$ and for (c) $\tau =13.8$} \label{Fig:k=1_a=0.05_topology}
\end{figure}

\noindent
{\bf Case 2. $\alpha=0.05\,( < \alpha_c)$} \footnote{ This is equivalent to the case $P<P_c$.
For $\varphi=0.5$ and $P=0.01$, the relation \eqref{critical_P_T} give $\alpha_c = (0.72 \pi)^{-1}\approx 0.44$. If  $\alpha=0.05$, then the critical pressure is  $P_c= (3.6\pi)^{-1}\approx 0.088$, so our choice  $P=0.001$ satisfies $P< P_c$.
Investigating $\alpha$ below for above $\alpha_c$ is thus equivalent to investigating  $P$ below or above its  critical value~$P_c$.}\\
Turning on the Gauss-Bonnet coupling $\alpha$ changes the topological properties compared to the case  $\alpha=0$. As the red curve in \autoref{k=1_tau_r_low_potential} suggests, there are two extremal points $\tau_a$ and $\tau_b$, with $\tau_a < \tau_b$. Depending on the range of $\tau$, the zero points differently appear as listed in \autoref{table:summary}: 
\begin{itemize}
    \item When $\tau < \tau_a$: There is one zero point, ($ZP_1$), as shown in \autoref{Plot_a}. To illustrate, we set $\varphi=0.5$, $k=1$, $P=0.01$ and $\tau=4.8$, and we find the zero point located at $(r_{c1},\Theta)=(15.1,\pi/2)$ in the $r_+ - \Theta$ plane, with corresponding winding number $\omega = +1$. This corresponds to the locally stable large black hole branch as can be seen in \autoref{k=1_tau_r_low_potential}. 
    \item When $\tau_a < \tau <\tau_b$: Three zero points, $(ZP_2, ZP_3, ZP_4)$, appear, as shown in \autoref{Plot_b}. To illustrate, we set $\tau = 8.6$ with the same other parameters, we find the zero points located at $\Theta=\pi/2$ and $(r_{c2}=0.127,r_{c3}=0.89, r_{c4}=7.7)$, and the corresponding winding numbers  $w= +1, -1,+1$, respectively. The new zero points $ZP_2$ and $ZP_3$ correspond to the locally stable small black hole and unstable intermediate black hole branches as can be seen in \autoref{k=1_tau_r_low_potential}.
    \item  When $\tau >\tau_b$: Two zero points $ZP_3$ and $ZP_4$ collapse into one zero point $ZP_5$. To illustrate, we set $\tau=13.8$ and find that the zero  point, $ZP_5$, is located at {$(r_{c5},\Theta)=(0.072,\pi/2)$}, with winding number $\omega = +1$,  corresponding to the locally stable small black hole branch as can be seen in \autoref{k=1_tau_r_low_potential}. The total winding number is $W = 1$, which is the same with the canonical ensemble. 
\end{itemize}

 \begin{table}[ht]
	\centering{\
		\begin{tabular}{|c |c |c|c|c |c|c|}
			\hline \hline 
	Case $k=1$& \#   & Winding & Topological No.\\
		$\alpha=0.05,\; \varphi=0.5$& of ZPs &  No. $(w_i)$ & $Q_t\equiv \sum w_i$ \\ \hline
			\multirow{3}{8em}{Case 1: $\tau < \tau_a$} &  &  &   \\ 
		                                        	& 1 & $w|_{\text{ZP}_1} = 1$ & $1$  \\ 
			                                        &  &  &   \\ \hline
             \multirow{3}{9em}{Case 2: $\tau_a< \tau < \tau_b$} &  & $w|_{\text{ZP}_2} = +1$ &   \\ 
			                                        & 3 & $w|_{\text{ZP}_3} = -1$ & $1$  \\ 
			                                        &  & $w|_{\text{ZP}_4} = +1$ &   \\ \hline
			 \multirow{3}{8em}{Case 3: $\tau > \tau_b $}&  &  &   \\ 
		                                        	& 1 & $w|_{\text{ZP}_5} = 1$ & $1$  \\
                                            &  &  &   \\ \hline
		\end{tabular}
	\caption{ The zero points for $\alpha=0.05$ and pressure less than the critical value, where $\tau_a = 5.5$ and $\tau_ b= 13.5$.}
	\label{table:summary}}
\end{table}

\begin{figure}[h]
\begin{minipage}{0.6\linewidth}
    \centering
   \includegraphics[width=1\textwidth]{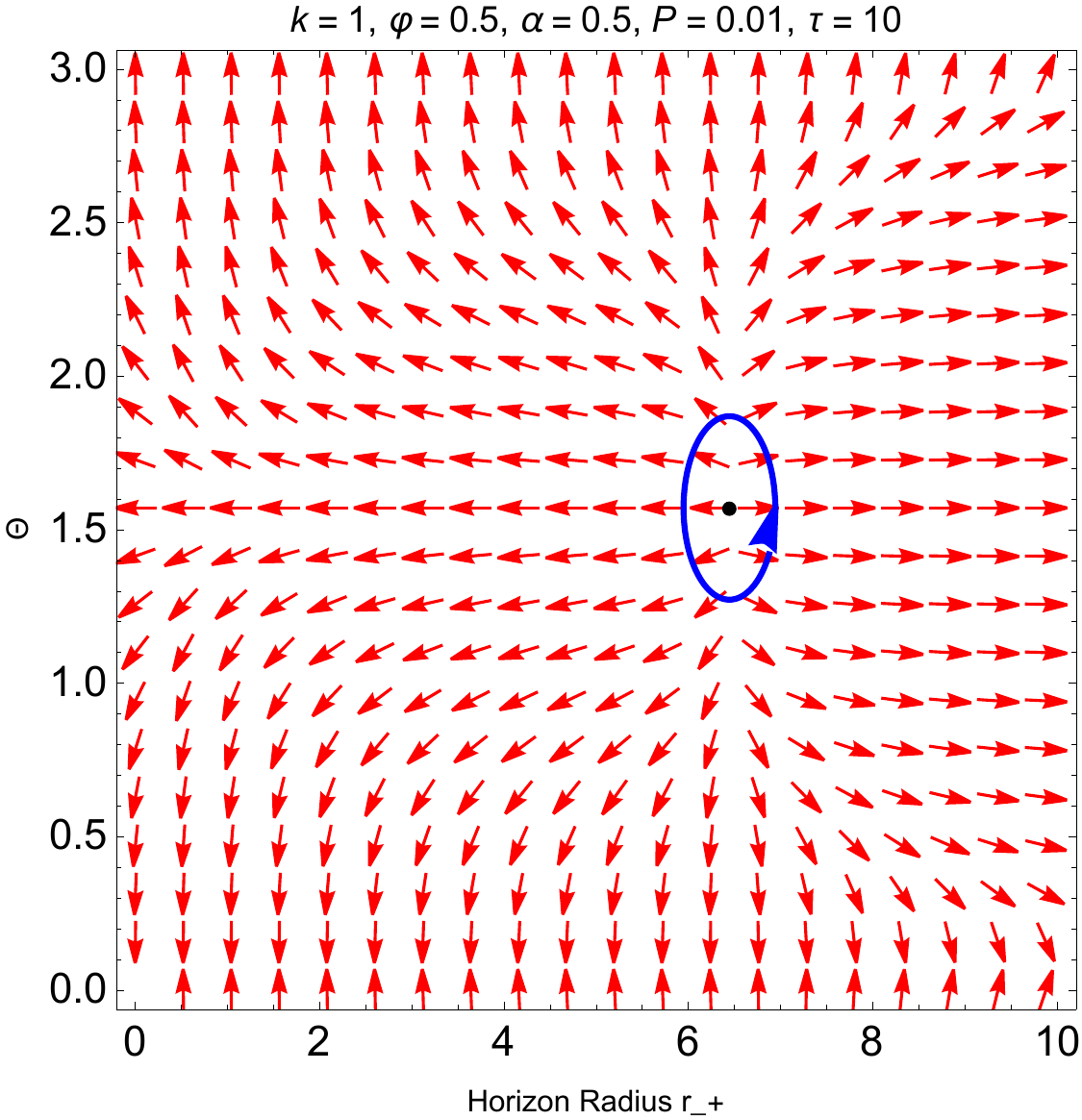}   
\end{minipage}
  \caption{\footnotesize  The unit vectors ($n^1,n^2$) in $\Theta$ vs $r_+$ plane, for coupling $\alpha=0.5$ and pressure $P=0.01$ higher than the critical value. }
\label{Fig:k=1_a=0.5_topology}
\end{figure}

\noindent
 {\bf Case 3: $\alpha=0.5 \,( > \alpha_c)$}
 \\
 For $P=0.01$ and $\varphi=0.5$, this corresponds to $\alpha >\alpha_c$ and, equivalently, $P>P_c$. In this case, the unstable intermediate region disappears, and the $\tau(r_+)$ curve shows only one stable black hole branch, as depicted by the blue curve in \autoref{k=1_tau_r_low_potential}. For $\tau=10$, we find that the zero point is located at $(r_{c}, \Theta) = (6.45, \pi/2)$, as shown in \autoref{Fig:k=1_a=0.5_topology}. The winding number for this black hole is $\omega = +1$, resulting in a total winding  number $W= +1$.

\subsubsection{$k=-1$}\label{Sec:k=-1_topology}
\begin{figure}
\begin{minipage}{0.28\textwidth}
    \centering
    \includegraphics[width=1\textwidth]{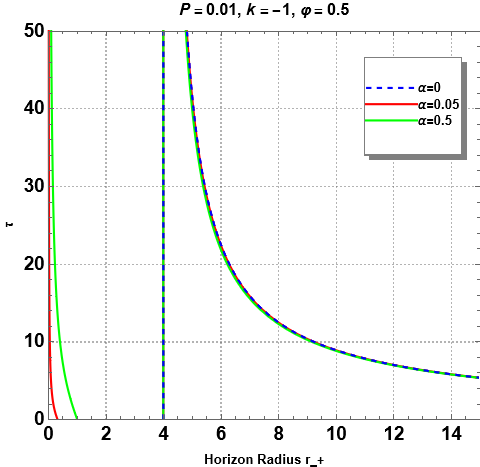}
    \subcaption{}
    \label{k=-1_tau_r_low_potential}
\end{minipage}
\hspace{0.1cm}
\begin{minipage}{0.28\textwidth}
    \centering
   \includegraphics[width=1\textwidth]{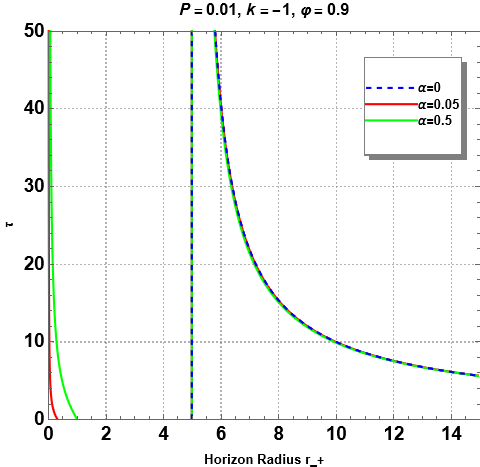}
    \subcaption{}  
      \label{k=-1_tau_r_high_potential}
\end{minipage}
\hspace{0.1cm}
\begin{minipage}{0.28\textwidth} 
    \centering
    \includegraphics[width=1\textwidth]{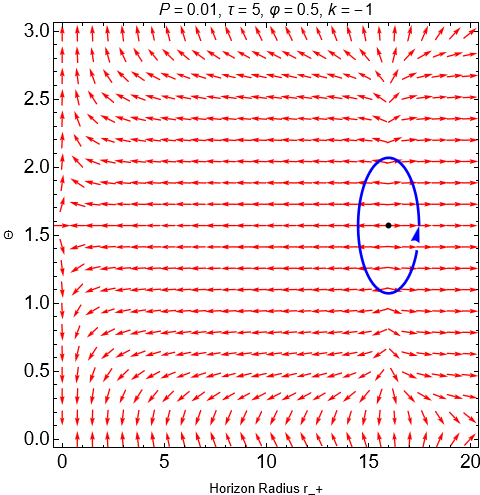}
      \subcaption{}
       \label{k=-1_zero_point}
\end{minipage} 
   \caption{\footnotesize The zero points of $\phi^{r_+}$ in $\tau$ vs $r_+$ plane for hyperbolic charged  AdS black hole in grand canonical ensemble with  $\tau=10$ and pressure $P=0.01$. Here for (a) $\varphi=0.5$, for (b) $\varphi=0.9$  (c) The unit vectors ($n^1,n^2$) in $\Theta$ vs $r_+$ plane for hyperbolic charged Gauss-Bonnet AdS black hole in grand canonical ensemble with coupling $\alpha=0.05$, $\tau=5$.} \label{Fig:k=-1_topology}
\end{figure}

In this case, the solution, \eqref{tausolution},  of the zero point of the vectors,\eqref{vector_defects} and \eqref{vector_phitheta}, becomes 
\begin{align}
  \tau =\frac{6 \pi  \left(r_+^2-2 \alpha \right)}{r_+ \left(8 \pi  P r_+^2-4 \varphi ^2-3\right)}\,.
\end{align}
From \eqref{enon}, we recall that the hyperbolic horizon radius must be $r_+^2 \geq 6\alpha$. Within this range, we observe that irrespective of the GB parameter $\alpha$ and potential $\varphi$, we get one stable black hole branch.  In \autoref{Fig:k=-1_topology}, we illustrate examples for various value of $\alpha$ for the choice of $\varphi =0.5$ and for the choice of $\varphi= 0.9 > \varphi_c^{\max}$. 
Correspondingly we find one zero point in $r_+ - \Theta$ plane as illustrated in \autoref{k=-1_zero_point}, where we choose, as an example, $\alpha=0.05$, $\varphi=0.5, P=0.01$ and $\tau=5$.  Further, we find the winding number $+1$, and hence the topological number is $+1$. This topological number is unchanged also in the change of pressure.  Therefore, in the hyperbolic-RN GB AdS black holes in the grand canonical ensemble have only one stable black hole solution.  

\section{Stability and Phase transition of RN GB Topological Black Holes}
\label{Sec:Stability_Specific_Heat}
\noindent
In this section, we explore the stability and phase transitions of charged Gauss-Bonnet black holes in the grand canonical ensemble by applying the laws of black hole thermodynamics. In \autoref{local_stability}, we will focus on the local stability of black holes for all three horizon geometry by determining the sign of specific heat. A key criterion for local thermodynamical stability is the positivity of specific heat capacities. In \autoref{phase_transition}, we will delve into the phase structure of black holes, exploring the conditions under which phase transitions occur and their physical significance. Together, these sections provide a comprehensive framework for understanding the thermodynamic nature of black holes.

\noindent

\noindent

\subsection{Local Stability}\label{local_stability}
The specific heat plays a crucial role in determining the local thermodynamic stability of a system. A positive specific heat indicates local stability, while a negative specific heat signals local instability due to perturbations in the system's parameters. In this subsection, we analyze the specific heat for several thermodynamic systems and verify whether branches with winding numbers of $(+1)$ or $(-1)$ correspond to positive or negative specific heat, respectively.

For a fixed potential in the grand canonical ensemble, the specific heat can be expressed as:
\begin{align}\label{Cvarphi}
     C_\varphi &= T \left( \frac{\partial S}{\partial T} \right)_{P,\varphi} = T \left(  \frac{\partial S}{\partial r_+}\right)_{P,\varphi} \left(  \frac{\partial r_+}{\partial T}\right)_{P,\varphi},
\end{align}
Notably, the sign of the specific heat $C_\varphi$ only depends on the sign of  $(\partial_{r_+} T)_{P, \varphi}$. 
Since, for the black hole entropy \eqref{S}
\begin{align}\label{S_r}
  \left( \frac{\partial S}{\partial r_+}\right)_{P,\varphi} = \frac{3 \pi ^2 \left(r_+^2+ 2 \alpha  k\right)}{2 } > 0\,,
\end{align}
 remains positive under the constraint \eqref{enon}.

\noindent
We obtain the expression of specific heat of an RN-GB AdS black hole in the grand canonical ensemble, by substituting the temperature \eqref{Tgrand} and \eqref{S_r} into \eqref{Cvarphi},
    \begin{align}
        C_\varphi 
       & =  -\frac{3}{2} \pi^2 r_+ \left(r_+^2 + 2 k \alpha \right)^2 \left(3 k+8 P \pi r_+^2 - 4  \varphi ^2\right) \nonumber
        \\&~~~~ \times  \Bigl( 8  k \alpha \varphi^2-4 r_+^2 (2 \pi P r_+^2 +  \varphi ^2) -6 k^2  \alpha 
        \\
        &~~~~~~~~~~~ + k r_+^2 (3-48 \pi  \alpha  P)\Bigr)^{-1} \,.\nonumber
      \end{align}
This formulation allows for the analysis of the thermodynamic stability of RN-GB AdS black holes with $k = \pm 1,0$ within the grand canonical ensemble.

\subsubsection{$k=0$}\label{k=0_local_stability}
We reiterate that in the $k=0$ case, the dependence on the Gauss-Bonnet coupling $\alpha$ vanishes. As a result, the thermodynamic properties of RN-GB AdS black holes are identical to those of RN AdS black holes.
In grand canonical ensemble, the specific heat is given as
\begin{align}\label{Ck=0}
   C_\varphi= \frac{3 \pi ^2 r_+^3 \left(2 \pi  P r_+^2-\varphi ^2\right)}{2 \left(2 \pi  P r_+^2+\varphi ^2\right)}\,.
\end{align}

Note from \eqref{cons_temp} that the condition for the non-negative temperature restricts the allowed value of the radius as
\begin{align}\label{rt0}
    r_+^2 \geq \frac{\varphi^2}{2 \pi P}= r_\text{T=0}^2\,.
\end{align}
This implies that the specific heat \eqref{Ck=0} is always non-negative for all allowed value of $r_+$. 
Therefore, we conclude that the RN-GB AdS black holes with a planar horizon $(k = 0)$ are stable. This is in accordance with the implication from the topological analysis in \autoref{Sec:k=0_topology}, where we find single zero point with winding number $w= +1$.

\begin{figure}[ht]
\begin{minipage}{0.6\linewidth}
    \centering
    \includegraphics[width=1\textwidth]{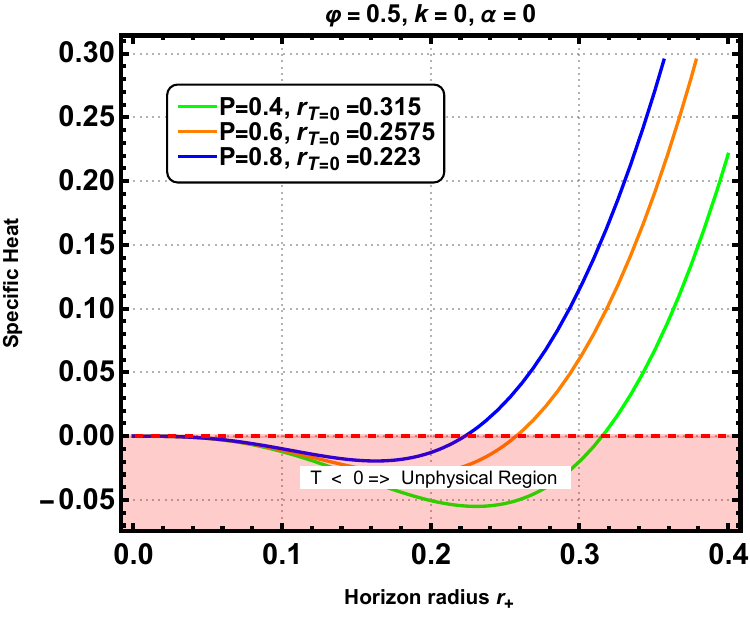}
    \label{k=0_cp_r_low_pressure}
\end{minipage}  
\caption{\footnotesize Specific Heat vs Horizon radius for $k=0$ with  for different parameters. }
	\label{Fig:k=0_cp_r}
\end{figure}

In \autoref{Fig:k=0_cp_r}, we depict the specific heat against horizon radius for various values of the pressure. The red-shaded area in Fig\eqref{Fig:k=0_cp_r} represents the unphysical solutions with negative temperature. Within the physical region, we note that the specific heat drops to zero at the minimum horizon radius $r_+= r_\text{T=0}$ (indicated by the red dotted line in Fig\eqref{Fig:k=0_cp_r}), and remains positive for larger radii ($r_+ > r_\text{T=0}$).

\subsubsection{$k=1$}\label{k=1_local_stability}
 The specific heat of spherical RN-GB AdS black hole $(k=1)$ in the grand canonical ensemble is given as
 \begin{align}
     C_\varphi = \frac{3 \pi ^2 r_+ \left( r_+^2 + 2 \alpha\right)^2 \left(8 \pi  P r_+^2-4 \varphi ^2+3\right)}{2 \left(\alpha  \left(6-8 \varphi ^2\right)+8 \pi  P r_+^4+r_+^2 \left(48 \pi  \alpha  P+4 \varphi ^2-3\right)\right)}\,.
 \end{align}

\begin{figure}[ht]
\begin{minipage}{0.6\linewidth}
    \centering
    \includegraphics[width=1\textwidth]{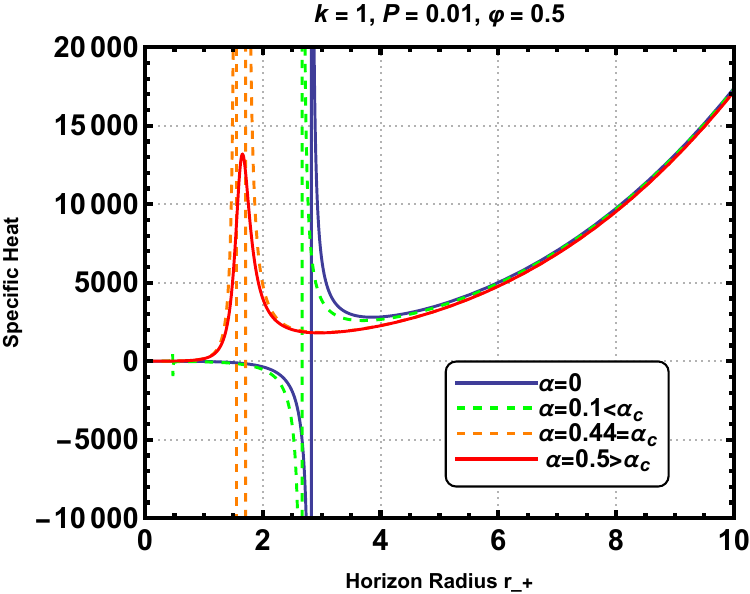}
     \subcaption{}
	\label{c1}
\end{minipage}
\hspace{0.1cm}
\begin{minipage}{0.6\linewidth} 
    \centering
\includegraphics[width=1\textwidth]{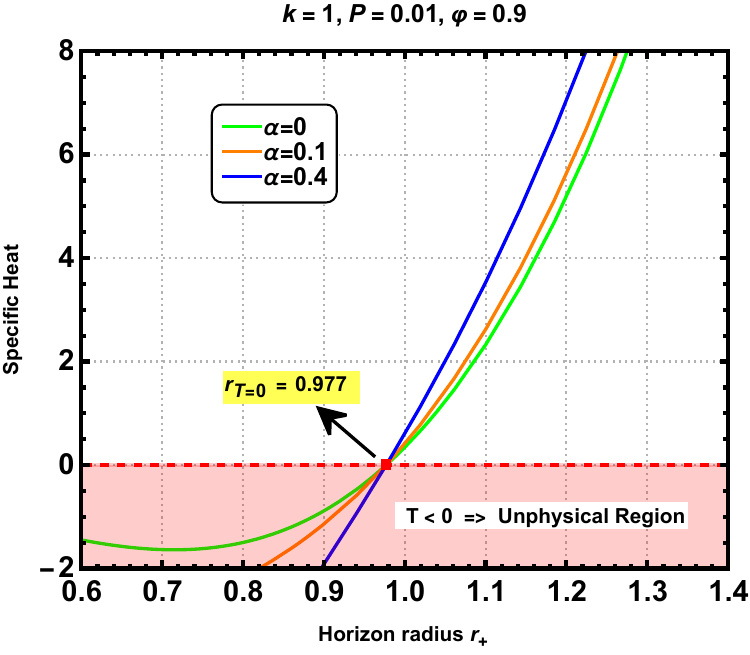}
 \subcaption{}
	\label{c2}
\end{minipage}  
   \caption{\footnotesize In the grand canonical ensemble for $k=1$ and pressure $P=0.01$: Specific Heat $C_\varphi$ versus Horizon radius $r_+$ for various potential values (a) $ \varphi=0.5 <\varphi_c^{\max}$ and (b) $\varphi= 0.9 > \varphi_c^{\max}$ with Gauss-Bonnet coupling ($\alpha=0, \alpha=0.1, \alpha=0.44$ and $\alpha=0.5$). }\label{Fig:k=1_cp_r_low_pressure}
\end{figure}

\noindent
In  Fig\ref{Fig:k=1_cp_r_low_pressure}(a) and Fig\ref{Fig:k=1_cp_r_low_pressure}(b), we present the $C_\varphi - r_+$ plot for two values of potential $\varphi =0.5 < \varphi_c^{\max} $ and $\varphi=0.9 > \varphi_c^{\max} $ respectively, with various choices of Gauss-Bonnet coupling at fixed pressure $P=0.01$. For an RN-GB black hole with a spherical horizon, all values of horizon radius are allowed as long as the black hole potential is $\varphi <\sqrt{3}/2$. However, for the potential $\varphi > \varphi_c^{\max}$, the temperature positivity imposes a constraint \eqref{cons_temp} on the horizon radius. The allowed value of the horizon radius is 
 \begin{align}\label{rt1}
     r_+^2 \geq \frac{4\varphi^2 -3}{8 \pi P} = r_\text{T=0}^2
 \end{align}
 As can be seen in Fig\ref{Fig:k=1_cp_r_low_pressure}(a), for $\varphi=0.5 < \varphi_c^{\max}$, we observe three black hole branches, corresponding to intermediate unstable, small, and large stable black holes, when the Gauss-Bonnet coupling $\alpha$ is below the critical value $\alpha_c =0.44$ which is determined by the relation~\eqref{critical_P_T}. 
 These were characterized by the winding numbers $w= +1, -1, +1$ respectively as described in \autoref{Black_Holes_Defects}. 
 As $\alpha$ exceeds this critical value ($\alpha=0.5>\alpha_c$), the unstable intermediate black hole branch disappears smoothly connecting stable small and large black hole branches.

\noindent
However, a distinct behavior emerges when the potential surpasses its maximum critical value $\varphi > \varphi_c^{\max}=\sqrt{3}/2$. In this regime, the black hole solutions exhibit a single branch irrespective of the value of GB coupling $\alpha$. Fig\eqref{Fig:k=1_cp_r_low_pressure}(b) shows the unphysical solutions with negative temperature in the red-shaded region.  As seen by the red dotted line in Fig\eqref{Fig:k=1_cp_r_low_pressure}(b), the specific heat within the physical area decreases to zero at the horizon radius $r_+= r_\text{T=0}$ and stays positive at higher radii ($r_+ > r_\text{T=0}$).

\noindent
Thus, we deduce that the spherical RN-GB AdS black holes $(k = 1)$ are locally thermodynamically stable.

 \subsubsection{$k=-1$}\label{k=-1_local_stability}

 \noindent
  The specific heat of hyperbolic RN-GB AdS black hole for the case $k=-1$ in the grand canonical ensemble is given as
\begin{align}
   C_\varphi= \frac{3 \pi ^2 r_+ \left(r_+^2-2 \alpha \right)^2 \left(8 \pi  P r_+^2-4 \varphi ^2-3\right)}{2 \left(8 \pi  P r_+^4+r_+^2 \left(4 \varphi ^2 -48 \pi  \alpha  P+3\right) + 2 \alpha  \left(4 \varphi ^2+3\right)\right)}\,.
\end{align}
 \begin{figure}[ht]
\begin{minipage}{0.6\linewidth}
    \centering
    \includegraphics[width=1\textwidth]{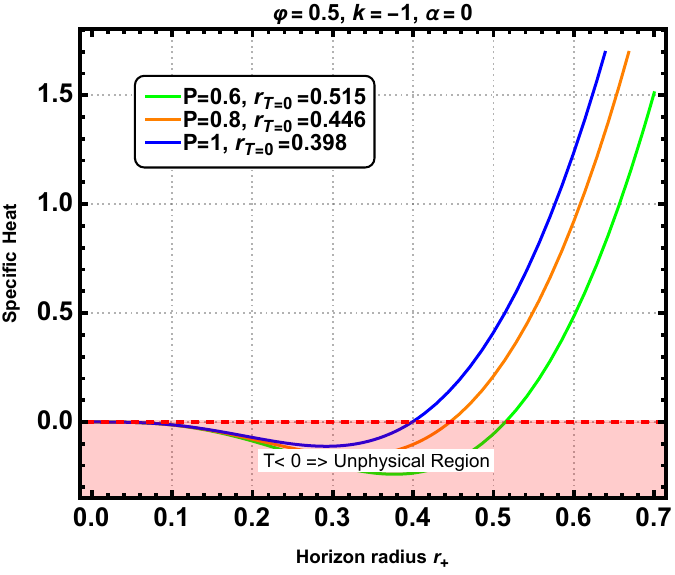}
     \subcaption{}
\end{minipage}
\hspace{0.1cm}
\begin{minipage}{0.6\linewidth} 
    \centering
    \includegraphics[width=1\textwidth]{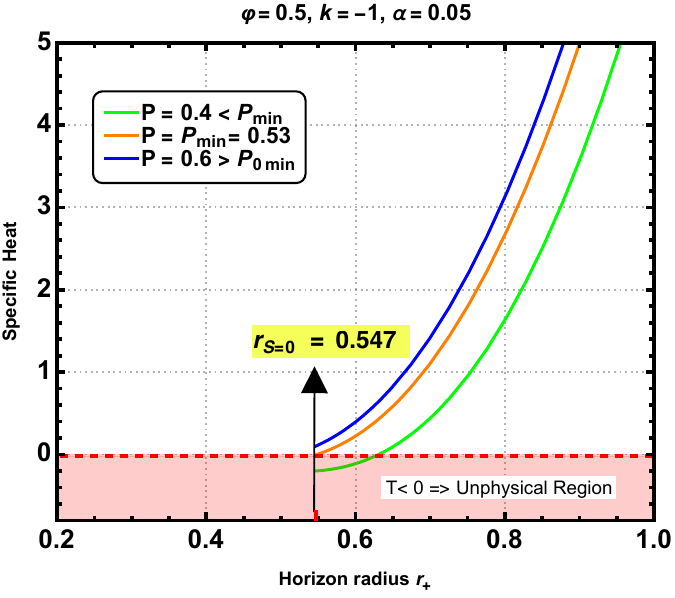}
     \subcaption{}
\end{minipage}  
\hspace{0.1cm}
\begin{minipage}{0.6\linewidth} 
    \centering
    \includegraphics[width=1\textwidth]{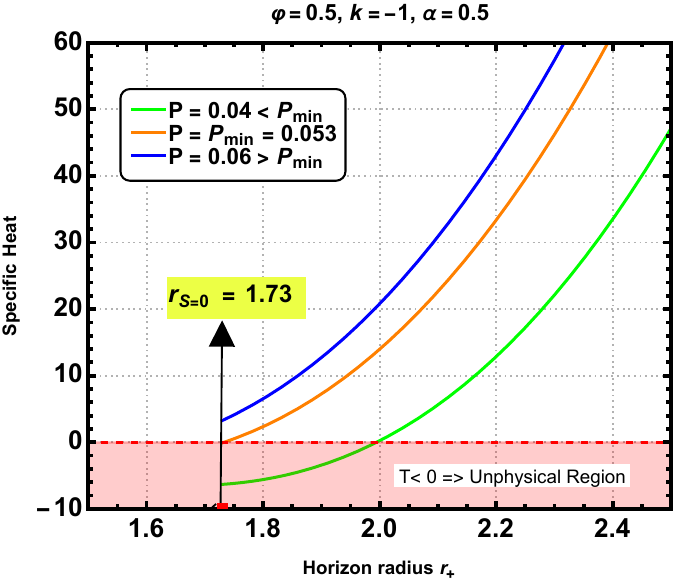}
     \subcaption{}
\end{minipage} 
\caption{\footnotesize Specific Heat vs Horizon radius in the grand canonical ensemble for $k=-1$ and $P=0.01$ for potential values. Here (a) $\alpha=0$, (b) $\alpha=0.05$, and (c) $\alpha=0.5$. }
	\label{Fig:k=-1_cp_r_low_pressure}
\end{figure}

\noindent
The horizon radius of this system is constrained by non-negative temperature \eqref{cons_temp} and non-negative definiteness of entropy \eqref{enon}. 
\begin{align}
  r_+ \geq \sqrt{\frac{3+4 \varphi^2}{8 \pi P}}= r_\text{T=0},\quad \; \quad r_+ \geq \sqrt{6\alpha} = r_\text{S=0}
\end{align}
In Fig\eqref{Fig:k=-1_cp_r_low_pressure}, we illustrate the specific heat vs horizon radius,  for potential $\varphi=0.5$ and various pressure values. In Fig\eqref{Fig:k=-1_cp_r_low_pressure}(a), we plot the specific heat of the hyperbolic RN AdS black hole $(\alpha=0)$ for the entire range of horizon radius.  

When the Gauss-Bonnet coupling is activated, hyperbolic black holes with horizon radii $r_+ < \sqrt{6\alpha}$ become non-physical due to non-negative definiteness of the entropy \eqref{enon}, so we omit them in Fig\eqref{Fig:k=-1_cp_r_low_pressure}(b) and Fig\eqref{Fig:k=-1_cp_r_low_pressure}(c) for coupling  $\alpha=0.05$ and $\alpha=0.5$, respectively. Furthermore, temperature positivity constrains the physical region to $(r_+\geq r_0)$.
The pressure $P_\text{min}$ corresponding to $r_\text{T=0} = r_\text{S=0}$ is given by
\begin{align}
    P_\text{min} = \frac{3+4 \varphi^2}{48 \pi \alpha}
\end{align}
 The red-shaded region in Fig\eqref{Fig:k=-1_cp_r_low_pressure} represents the forbidden area due to negative temperatures. The specific heat remains positive in the physical region, indicating local thermodynamic stability of hyperbolic RN-GB AdS black holes. This is consistent with \autoref{Sec:k=-1_topology}, where the branch with a positive winding number $+1$ is locally thermodynamically stable. 

\subsection{Phase Transition }\label{phase_transition}
 
In this subsection, we investigate the effect of the Gauss-Bonnet coupling on the phase structure of the black holes within the grand canonical ensemble framework. To elucidate the phase transition, we construct $G-T$ diagrams, where the Gibbs free energy $(G)$ is plotted against temperature $(T)$ while maintaining the other parameters constant. A thermodynamically stable state is determined by the global minimum of the Gibbs free energy $(G)$, which considers both the black hole configurations and the background spacetime. Notably, the background spacetime serves as a reference point, which is characterized by zero Gibbs free energy. Consequently, a negative G indicates that the corresponding phase is thermodynamically favored over the background spacetime, signifying a stable black hole configuration. We will study the nature of phase transition in RN-GB AdS black holes with $k= \pm 1, 0$ within the grand canonical ensemble.

 \subsubsection{$k=0$}
\begin{figure}[ht]
\begin{minipage}{0.6\linewidth}
    \centering
    \includegraphics[width=1\textwidth]{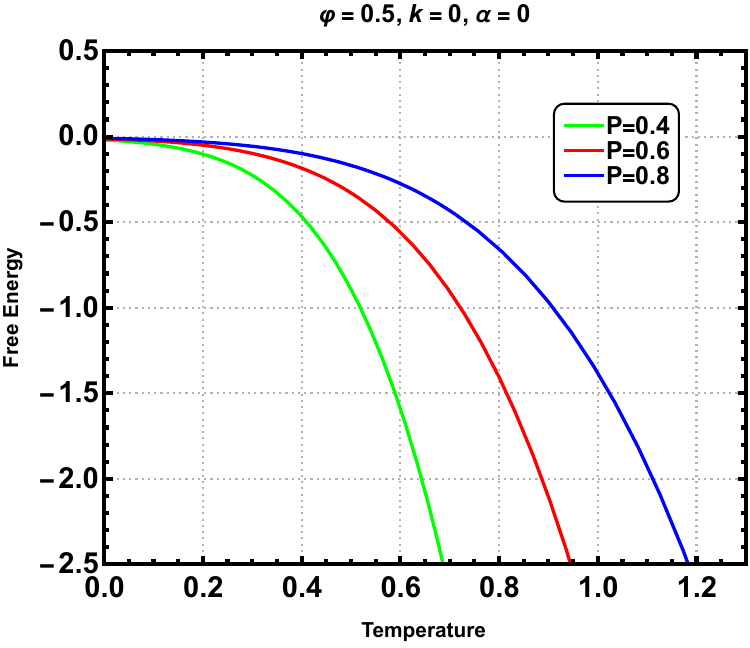}
\end{minipage}
\caption{\footnotesize Phase transition for $k=0$ with  for different parameters. }
	\label{fig:FT_k=0}
\end{figure}

 The temperature $T$ and on-shell Gibbs free energy $G$ for $k=0$ case is given as   
\begin{align}
   T =  \frac{4 P r_+}{3}-\frac{2 \varphi ^2}{3 \pi  r_+}, \quad {G} = -\frac{1}{6} \pi  r_+^2 \left(\pi  P r_+^2+\varphi ^2\right)\,.
\end{align}
The \autoref{fig:FT_k=0}, illustrates the general behavior of the Gibbs free energy, as depicted in $G-T$ diagrams for various pressures. Notably, the Gibbs free energy of the RN-GB AdS black hole with $k=0$ remains consistently negative. This indicates that AdS black hole v with the planar event horizon is thermodynamically favored over the AdS background spacetime phase, hence regarded as the globally stable state.

    \subsubsection{$k=1$}
The temperature $T$ and on-shell Gibbs free energy $G$ for spherical RN-GB black holes are given as  
\begin{eqnarray}
    T &= &\frac{8 \pi  P r_+^3-4 r_+ \varphi ^2+3 r_+}{12 \pi  \alpha +6 \pi r_+^2}\,,\\
   \label{G_for_k=1} {G} &=& \frac{\pi  \left(18 \alpha ^2-4 \pi  P r_+^6 - r_+^4 \left(72 \pi  \alpha  P-3\right)-9 \alpha  r_+^2 \right)}{24 \left(2 \alpha +r_+^2\right)}
   \nonumber \\
    && -\frac{\varphi^2 \pi r_+^2 (    r_+^2  - 6  \alpha )     }{6 \left(2 \alpha +r_+^2\right)}\,.
\end{eqnarray}

\begin{figure}[ht]
\begin{minipage}{0.6\linewidth}
    \centering
    \includegraphics[width=1\textwidth]{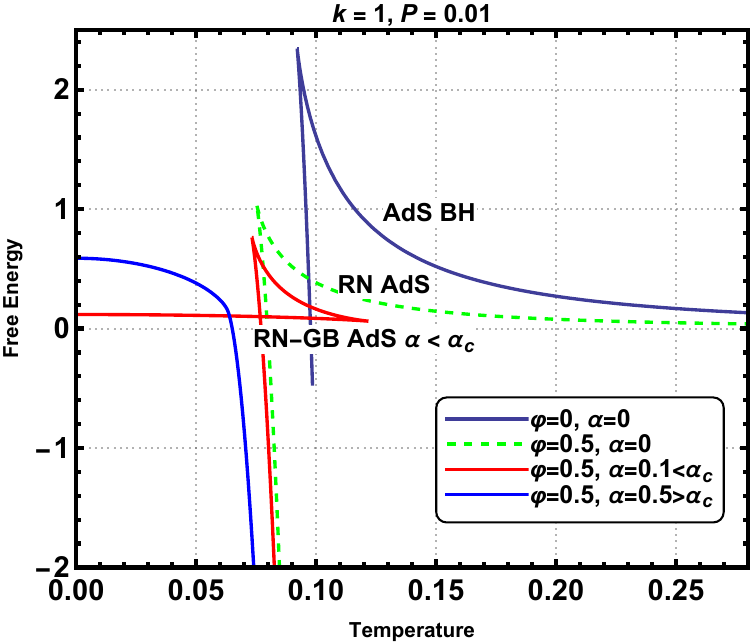}
    \subcaption[a]{}
\end{minipage}
\hspace{0.1cm}
\begin{minipage}{0.6\linewidth} 
    \centering
    \includegraphics[width=1\textwidth]{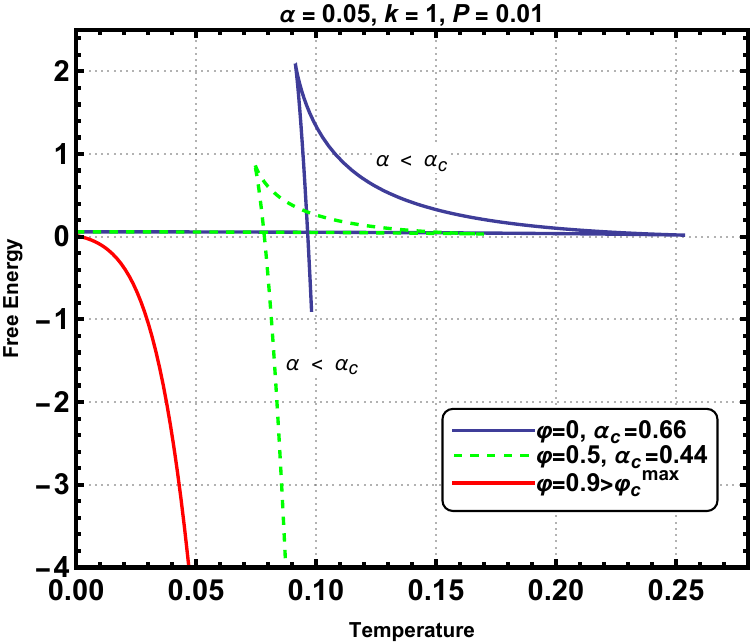}
    \subcaption[b]{}
\end{minipage}  
\hspace{0.1cm}
\begin{minipage}{0.6\linewidth} 
    \centering
    \includegraphics[width=1\textwidth]{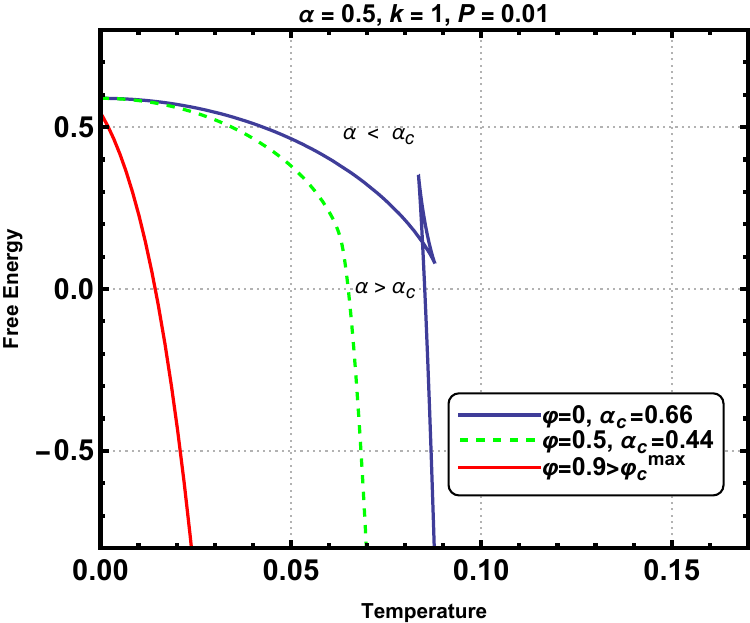}
    \subcaption[]{}
\end{minipage} 
\caption{\footnotesize Phase transition in the grand canonical ensemble for $k=1$ and $P=0.01$ for potential values. Here (a) $\alpha=0$, (b) $\alpha=0.05$, and (c) $\alpha=0.5$. }
	\label{fig:k=1_FT_low_pressure}
\end{figure}

\noindent
Fig\ref{fig:k=1_FT_low_pressure}(a) illustrates the $G-T$ relation for various configurations with spherical horizon $(k=1)$, including AdS $(\alpha =0\,, \varphi = 0)$, RN AdS $(\alpha = 0 \,, \varphi\neq 0)$, and RN-GB AdS $(\alpha \neq 0 \,, \varphi \neq 0)$ black holes.
 The Fig \ref{fig:k=1_FT_low_pressure}(b) and the Fig \ref{fig:k=1_FT_low_pressure}(c) both illustrate the effect of Gauss-Bonnet coupling $\alpha$ in $G-T$ diagram of various electric potentials for $\alpha=0.05$ and $\alpha = 0.5$, respectively.

 Notably, the Gauss-Bonnet coupling completely changes the phase structure. Unlike the AdS (blue) and RN-AdS (green) black hole phases shown in Fig \ref{fig:k=1_FT_low_pressure}(a), which have two distinct phases, GB-AdS black holes have one or three phases;  
 When the potential is in the range $0\leq \varphi \leq \varphi_c^{\max}$, a critical coupling threshold $\alpha_c$ exists, beyond which ($\alpha>\alpha_c$) there is only one black hole solution,  and below which $(\alpha<\alpha_c)$ there are three distinct black hole phases displaying the classical swallow-tail behavior. On the other hand, for potential $\varphi=0.9 > \varphi_c^{\max}$ we see one black hole phase for all $\alpha$ values, as shown by the red curves in Fig\ref{fig:k=1_FT_low_pressure}(b) and Fig\ref{fig:k=1_FT_low_pressure}(c). 

This analysis is consistent with the topology-thermodynamics analysis presented in Sec.[\ref{Sec:k=1_topology}]. For the potential $0\leq \varphi \leq \varphi_c^{\max}$, we observed three branches of black hole solutions for $\alpha<\alpha_c$, characterized by winding numbers $+1, -1, +1$. In contrast, for $\alpha>\alpha_c$, the system exhibits one black hole branch with winding numbers $+1$. Furthermore, for potential values exceeding the critical value $\varphi>\varphi_c^{\max}$, our analysis revealed a single black hole branch, as depicted in Fig.\ref{k=1_zero_point_high_potential} with  $w = +1$.

\noindent

The above analysis tells us that liquid-gas type phase transition is allowed even in the grand canonical ensemble in the presence of the Gauss-Bonnet coupling. 
However, we should note that the grand canonical ensemble, where the potential is fixed and the charge $Q$ is allowed to vary, also permits the Hawking-Page transition. This is in contrast to the canonical ensemble, where the charge is fixed, and a charged black hole solution cannot collapse into an uncharged AdS solution due to charge conservation.
Therefore, the actual phase transition is determined by which system has the minimum Gibbs free energy.

Fig.\ref{fig:k=1_FT_low_pressure} reveals that the thermal AdS vacuum phase is thermodynamically favored over locally stable small black holes corresponding to low temperatures. This is evident from the fact that the Gibbs free energy of the small black holes is positive $(G > 0)$, whereas the AdS solution has a vanishing Gibbs free energy $(G = 0)$ and due to the global stability of the AdS vacuum phase surpassing that of the small black hole, the system preferably undergoes the Hawking-Page phase transition.  
The system does go under a first-order
phase transition between a large black hole and a thermal AdS
vacuum at $G=0$ and Hawking-Page temperature $T_{HP}$. We can directly solve for $G = 0$ in \eqref{G_for_k=1}, to get the Hawking-Page temperature $T_{HP}$,  
   \begin{align}\label{THP}
    T_{HP} = \frac{6 \alpha -4 r_+^2 \varphi ^2+3 r_+^2}{2 \pi  r_+^3+36 \pi  \alpha r_+}\,.
\end{align} 
\noindent


\subsubsection{$k=-1$}
The temperature $T$ and on-shell Gibbs free energy $G$ for hyperbolic RN-GB black holes are given as  
\begin{eqnarray}
T &=& \frac{r_+ \left(8 \pi  P r_+^2-4 \varphi ^2-3\right)}{6 \pi  \left(r_+^2-2 \alpha \right)}\label{temp_hyperbolic}\,,\\
 {G} &=& -\frac{\pi  \left(18 \alpha ^2+4 \pi  P r_+^6 + r_+^4 \left(3-72 \pi  \alpha  P\right)+9 \alpha  r_+^2 \right)}{24 \left(r_+^2-2 \alpha \right)}\nonumber
 \\
 && -\frac{\pi \varphi^2 r_+^2 (  r_+^2  + 6 \alpha  )}{ 6 \left(r_+^2-2 \alpha \right)} \label{k=-1Gibbs}
\end{eqnarray}
 \begin{figure}[ht]
\begin{minipage}{0.6\linewidth}
    \centering
    \includegraphics[width=1\textwidth]{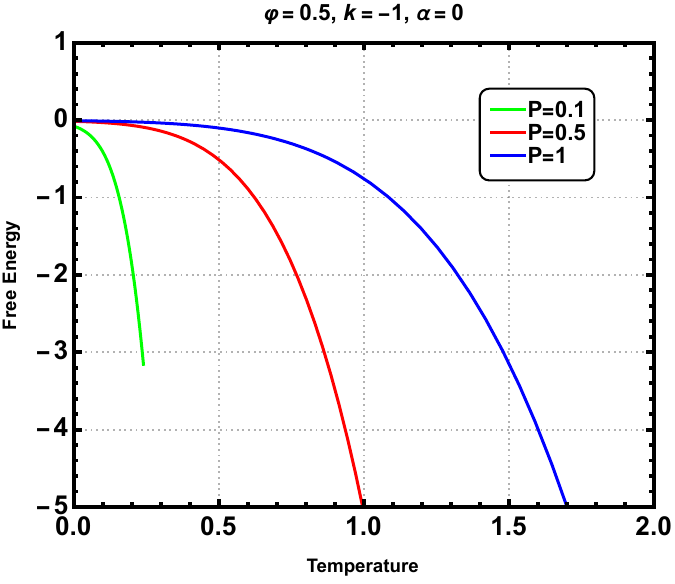}
     \subcaption{}
\end{minipage}
\hspace{0.1cm}
\begin{minipage}{0.6\linewidth} 
    \centering
    \includegraphics[width=1\textwidth]{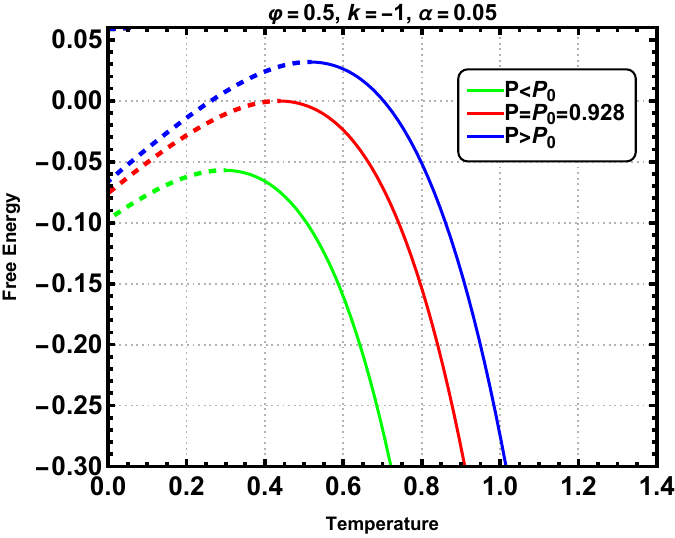}
     \subcaption{}
\end{minipage}  
\hspace{0.1cm}
\begin{minipage}{0.6\linewidth} 
    \centering
    \includegraphics[width=1\textwidth]{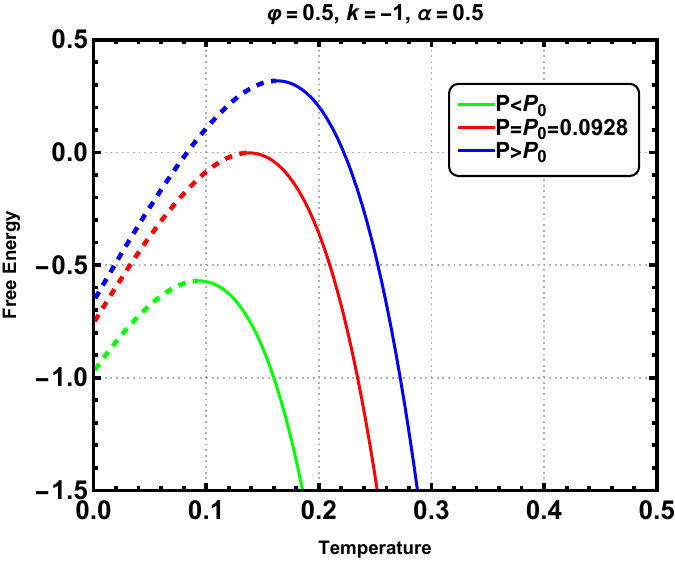}
     \subcaption{}
\end{minipage} 
\caption{\footnotesize Phase transition in the grand canonical ensemble for $k=-1$ and various values of pressure. Here (a) $\alpha=0$, (b) $\alpha=0.05$, and (c) $\alpha=0.5$. }
\label{fig:k=-1_FT_low_pressure}
\end{figure}

\noindent
Here, we recall that unlike RN-GB spherical $(k=1)$ and planar $(k=0)$  AdS black holes, the horizon radius of a hyperbolic RN-GB AdS black hole is bounded as $r_+ \geq \sqrt{6 \alpha}$ due to entropy positivity \eqref{enon}.

\noindent

Without Gauss-Bonnet coupling, the Fig\eqref{fig:k=-1_FT_low_pressure}(a), shows that 
$G$ is always negative for all values of pressure. Therefore, the single branch of the locally stable black hole solutions studied in the previous section all remain globally stable and there is no Hawking-Page phase transition. 

However, as we switch on the Gauss-Bonnet coupling, postive part of $G$ can appear. 
The Fig\eqref{fig:k=-1_FT_low_pressure}(b) and  Fig\eqref{fig:k=-1_FT_low_pressure}(c) plot the $G- T$ diagram for two different Gauss-Bonnet coupling, $\alpha = 0.05$ and $\alpha=0.5$ respectively, with fixed potential $\varphi=0.5$, where the dotted lines represents the forbidden region corresponding to $r_+^2 < 6\alpha \equiv r_{S=2}^2$~\footnote{In \cite{Cui:2021qpu}, the dotted lines were also considered, where there is additional Hawking-Page transition.}. The figures show that there exist a threshold pressure $P_0$ for given $\alpha$ and $\varphi$ such that for $P > P_0$, the Gibbs free energy becomes positive under a certain temperature $T_{HP}$ (green and red lines). At this temperature Hawking page transition occurs between massive black hole and massless black holes and AdS background spacetime (with zero Gibbs free energy).  
For $P<P_0$   Gibbs free energy is always negative (green line) and there is no Hawking-Page transition.

The Hawking Page temperature $T_{HP}$ is given by setting $G=0$. From \eqref{k=-1Gibbs}, we obtain the pressure as
\begin{align}
P_{HP} = - \frac{18 \alpha^2 +r_+^4 \left(3+4\varphi^2\right)+3\alpha r_+^2\left(3+8\varphi^2\right) }{4\pi r_+^4 \left(r_+^2 -2 \alpha\right)}
\end{align}
and substituting it into Eq.\eqref{temp_hyperbolic}, we get
\begin{align}
    T_{HP}= \frac{6\alpha - r_+^2 \left(3+4\varphi^2\right)}{2 \pi r_+ \left(r_+^2 - 18 \alpha\right)}\,.
\end{align}

The threshold pressure $P_0$ can be found by imposing additional condition $\partial_{r_+}P =0$. This is solve by $r_+ = \sqrt{6\alpha}$, which is within allowed range given in \eqref{enon}. This obtains the threshold pressure and the corresponding temperature $(P_0\,,T_0)$
\begin{align}
  P_0 = \frac{5 + 8 \varphi^2}{48 \pi \alpha},\;\;\quad T_0=  \frac{1 + 2 \varphi^2}{2 \pi \sqrt{6 \alpha}}
\end{align}

\begin{figure}[ht]
\begin{minipage}{0.6\linewidth} 
    \centering
    \includegraphics[width=1\textwidth]{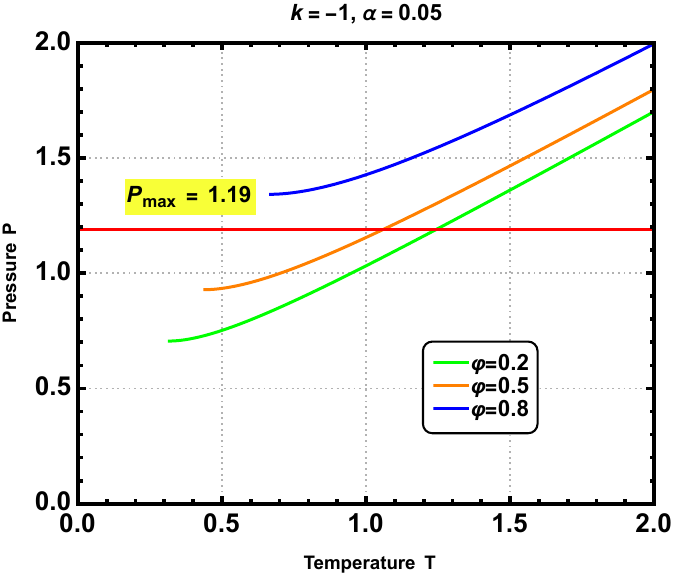}
     \subcaption{}
\end{minipage}  
\hspace{0.1cm}
\begin{minipage}{0.6\linewidth} 
    \centering
    \includegraphics[width=1\textwidth]{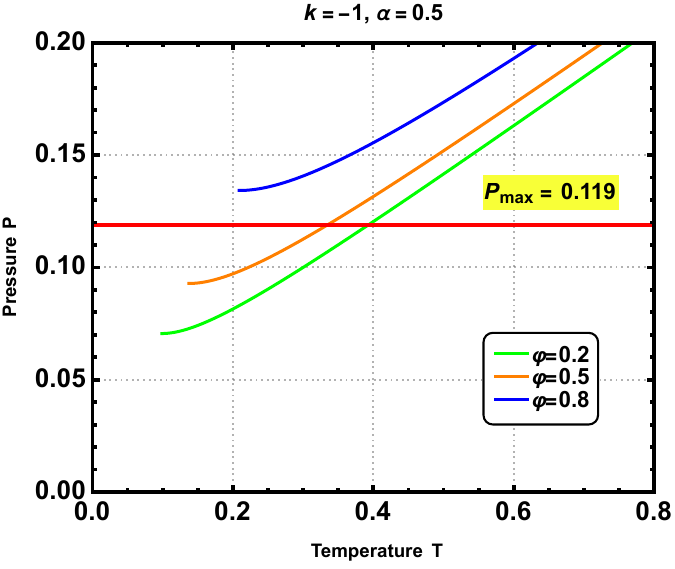}
     \subcaption{}
\end{minipage} 
\caption{\footnotesize Pressure vs Temperature plot for different potential values in the allowed region $r_+\geq \sqrt{6\alpha}$. Here (a) $\alpha=0.05$ and (b) $\alpha=0.5$ }
\label{fig:k=-1_PT}
\end{figure}

 Remember that the condition for the well-defined asymptotic vacuum \eqref{Pmax} restricts the pressure $P \leq P_\text{max}= {3}/{(16 \pi \alpha)}$. Therefore, the Hawking-Page phase transition can occur provided the pressure of the system lies in the range $P_0 < P \leq P_\text{max}$. If  $P_{\max}<P_0$, then there is no Hawking-Page transition. This means that  there is a maximum potential 
\begin{align}
    \varphi_\text{max}^\text{HP}= \frac{1}{\sqrt{2}} 
\end{align}                                                                    
 The Fig \ref{fig:k=-1_PT} shows the relation between the Hawking temperature and the pressure. (The forbidden regions corresponding to $r_+^2 < 6\alpha$ are omitted.) The figures show that when $\varphi > \varphi^{\text{HP}}_{\max}$, the pressure exceeds its allowed upper bound $P_{\max}$, and  thus the Hawking-Page transitions do disappear. 

\section{Conclusion}\label{conclusion}
We have presented a topological analysis for the thermodynamics of five-dimensional charged Gauss-Bonnet black holes in a grand canonical ensemble. As a complementary study, we have also presented the local stability given by the specific heat, which supports the implication of the topological analysis, and further presented the phase structure determined by the global stability, which is not implied by the topological analysis.

In cases, $k=-1$ and $k=0$, black holes with and without Gauss-Bonnet coupling have the same topological properties. For both cases, there are no critical points in the thermodynamic phase diagram, and there exists one locally stable black hole with a topological number $w=+1$. 

For the case $k=1$, the presence of the Gauss-Bonnet coupling $\alpha$ generates non-trivial thermodynamic structures in the grand canonical ensemble. 
For non-zero positive Gauss-Bonnet coupling $\alpha > 0$ within the potential range $0\leq \varphi \leq \varphi_c^{\max}$, the system exhibits the conventional critical point with topological charge $Q_t =-1$. In treatment of black hole solutions as defects, the topological classification is different after turning on the Gauss-Bonnet coupling, as the total winding number for the RN-GB AdS black hole is $W=+1$, while the one for the RN-AdS black hole is $W= 0$. This is the same topological class as the RN AdS and RN-GB AdS black holes in canonical ensemble \cite{Liu:2022aqt}. 

 For the potential $0\leq \varphi \leq \varphi_c^{\max}$, there are three branches of black hole solutions for $\alpha<\alpha_c$, with the corresponding winding numbers $w= +1, -1, +1$, and one black hole solution for $\alpha>\alpha_c$ with the corresponding winding numbers $w= +1$. 
 Furthermore, for potential $\varphi>\varphi_c^{\max}$, there is only one black hole branch with $w = +1$. All cases have the total winding number $W= +1$.  The branches with the winding number $w= +1$ and $w= -1$ correspond to the locally stable and unstable black hole solutions, respectively, and this is confirmed by the analysis through the specific heat. 

Although the topological analysis reveals the nature of critical points and the information on the locally stable and unstable phases, it does not  tell us about the global phase structure. The spherical $(k=1)$ RN-GB AdS black hole in the grand canonical ensemble locally allows the liquid-gas type phase transition. However, the global analysis using the Gibbs free energy shows that the thermal AdS vacuum phase is thermodynamically favored over the locally stable small black hole and the system preferably undergoes the Hawking-Page phase transition between the large black hole and a thermal AdS vacuum.  The hyperbolic black holes have only one black hole branch, however this also undergoes the Hawking-Page phase transition between massive black hole and AdS background.

There are several intriguing directions for future research.  Apart from the topological classification of various black hole systems, it is important to note that the current analysis  has been confined to the classical regime. 
Since the parameters considered are not limited to the thermodynamic regime, the quantum corrections to the thermodynamic quantities become significant  when the system moves beyond the thermodynamic limit \cite{Sen:2012kpz}. Notably, the emergence of three black hole phases with 1iquid-gas-type phase structures occurs at the finite value of the electric charge, situating them outside the thermodynamic regime. 
Furthermore, we should note that at very small temperatures, the classical analysis  breaks down \cite{Preskill:1991tb}.  As zero-temperature black holes are generically unstable, new approaches are necessary to explore their properties in the extremal limit. 
However, supersymmetric black holes, which posesses well-defined zero temperature statem, exhibit non-trivial phase structure even in this limit. Investigating these phases would be a particularly interesting direction for future research.  
\\

\vspace{-2mm}
\section*{Acknowledgements} 
\indent  This research was supported by an appointment to the  Young Scientists Training Program (M.M.) and JRG Program at the Asia Pacific Center for Theoretical Physics (APCTP) through the Science and Technology Promotion Fund and Lottery Fund of the Korean Government, by the Korean Local 
Governments - Gyeongsangbuk-do Province and Pohang City, and  by the National 
Research Foundation of Korea (NRF) grant funded by the Korean government (MSIT) (No. 2021R1F1A1048531). B.-H. Lee (RS-2024-00339204)
 and W. Lee (2022R1I1A1A01067336) and Center for Quantum SpaceTime (CQUeST )
 (Grant No. 2020R1A6A1A03047877) were
 supported by the Basic Science Research Program through the National Research Foundation of Korea is funded by the Ministry of Education.
We wish to thank Alfredo Gonz\' alez Lezcano, Augniva Ray 
for many interesting and useful discussions related to the topics discussed in this paper. B.-H.L appreciates the hospitality of APCTP and Korea Institute for Advanced Study.


\providecommand{\href}[2]{#2}\begingroup\raggedright\endgroup

\end{document}